\documentclass[]{emulateapj}
\usepackage{natbib}
\usepackage{float}
\usepackage{multirow}
\usepackage{threeparttable}
\usepackage{amsmath}	
\usepackage{amssymb}	
\usepackage{cases}

\newcommand{\kms}{km$\rm s^{-1}$}
\newcommand{\masyr}{\,mas\,$\rm yr^{-1}$}

\newcommand {\Msun}{{M$_{\odot}$}}

\def\max{{\rm max}}
\def\tot{{\rm tot}}
\def\tidal{{\rm tidal}}
\def\typ{{\rm typ}}

\newcommand{\Gaiaurl}{\href{https://gea.esac.esa.int/archive/documentation/GDR2/Gaia_archive/chap_datamodel/sec_dm_main_tables/ssec_dm_gaia_source.html}
{\url{https://gea.esac.esa.int/archive/documentation/GDR2/}}}

\PassOptionsToPackage{pdfpagelabels=false}{hyperref}
\PassOptionsToPackage{nonamebreak}{natbib}

\shorttitle{Separation Distributions of Ultra-wide Binaries}
\shortauthors{HJT, et al.}

\begin{document}

\title{The Separation Distribution of Ultra-Wide Binaries\\ across Galactic Populations}
\author{Hai-Jun Tian$^{1,2,3}$}
\author{Kareem El-Badry$^{3,4}$}
\author{Hans-Walter Rix$^{3}$}
\author{Andrew Gould$^{3}$}

\affil{$^{1}$China Three Gorges University, Yichang 443002, China}
\affil{$^{2}$Center for Astronomy and Space Sciences, China Three Gorges University, Yichang 443002, China}
\affil{$^{3}$Max Planck Institute for Astronomy, K\"onigstuhl 17, D-69117 Heidelberg, Germany}
\affil{$^{4}$Department of Astronomy and Theoretical Astrophysics Center, University of California Berkeley, Berkeley, CA 94720}

\begin{abstract}
We present an extensive and pure sample of ultra-wide binary stars with separations of $0.01 \lesssim s/{\rm pc} \lesssim 1$ in the solar neighborhood.
Using data from {\it Gaia} DR2, we define kinematic sub-populations via the systems' tangential velocities, i.e., disk-like ($v_{\perp, \rm tot}\le 40\,\rm km\,s^{-1}$), intermediate ($v_{\perp,\rm tot}=(40-85)\,\rm km\,s^{-1}$), and halo-like ($v_{\perp,\rm tot}\ge 85\,\rm km\,s^{-1}$) binaries, presuming that these velocity cuts represent a rough ordering in the binaries' age and metallicity.
Through stringent cuts on astrometric precision, we can obtain pure binary samples at such wide separations with thousands of binaries in each sample. For all three kinematic populations, the distribution of binary separations extends smoothly to 1\,pc, displaying neither strong truncation nor bimodality. 
Fitting a smoothly-broken power law for the separation distribution, we find its slope at $s=10^{2.5-4}$\,AU is the same for all sub-populations, $p(s)\propto s^{\gamma}$ with $\gamma \approx -1.54$. However, the logarithmic slope of $p(s)$ steepens at $s \gtrsim 10^4$\,AU, to $\gamma \gtrsim 2$. We find some evidences that the degree of steepening increases with the binaries' age, with a slope-change of only $\Delta\gamma\approx 0.5$ for disk-like stars, but $\Delta\gamma\gtrsim 1$ for halo-like stars. 
This trend is contrary to what might be expected if steepening at wide separations were due to gravitational perturbations by molecular clouds or stars, which would preferentially disrupt disk binaries.
If we were to interpret steepening at $s\gtrsim 10^4$\,AU as a consequence of disruption by MACHOs, we would have to invoke a MACHO population inconsistent with other constraints.  As a more plausible alternative, we propose a simple model to predict the separation distribution of wide binaries formed in dissolving star clusters. 
This model generically predicts $\gamma\simeq -1.5$ as observed, with steepening at larger separations due to the finite size of binaries' birth clusters. 

 \end{abstract}

\section{Introduction}\label{sect:intro}
It has long been recognized that wide binaries provide a powerful tracer of the Galactic gravitational potential on small scales \citep[e.g.][]{Bahcall1981, Bahcall1985} as well as useful constraints for studies of the star formation process \citep{van1968,Chaname2007,Moeckel2010,Kouwenhoven2010,Moeckel2011}. The orbits of wide binaries are so fragile (with escape velocities $v_{\rm esc}\lesssim $\kms) that they can be easily unbound by external gravitational perturbations arising from molecular clouds, stars, or compact objects \citep[e.g.][]{Retterer1982, Bahcall1985, Weinberg_1987, Jiang_2010, Allen2014ApJ}, or by internal perturbations due to evolution of the component stars \citep{Boersma_1961, Savedoff_1966, El-Badry2018a}. 

One of the most important applications of wide binaries is to constrain the mass of viable MAssive Compact Halo Objects (MACHOs), such as discrete black holes (BHs), which are a proposed dark matter candidate. Studies using wide binaries to constrain MACHOs have typically assumed a simple initial separation distribution (such as a smooth power law) and then interpreted deviations from it as arising from disruption of the widest binaries by MACHOs. 

A number of previous studies have used halo wide binaries -- which are expected to be less vulnerable to disruption by molecular clouds -- to constrain the plausible parameters of MACHOs. \citet[][hereafter CG04]{CG04} built a catalog of 1147 candidate wide binaries selected via common proper motion, from which they selected 801 binaries with disk-like orbits and 116 binaries with halo-like orbits. Fitting the two subsamples, they showed that the angular separation distributions could be well described by single power laws with logarithmic slopes of -1.67$\pm$0.07 for the disk binaries and -1.55$\pm$0.10 for the halo binaries, over the angular separation range $3.5''<\Delta\theta<900''$. In a companion paper, \citet[][hereafter Y04]{Y04} argued that the CG04 separation distribution constraints ruled out most of the previously plausible MACHO parameter space, since there was no strong evidence of a break in the separation distribution at the widest separations, and their simulations predicted a detectable break for MACHO masses $M\gtrsim 43\,M_{\odot}$.

When combined with constraints from microlensing surveys (which ruled out a dominant MACHO population with typical masses in the range of $10^{-7} \lesssim M/M_{\odot} \lesssim 30$; \citealt{Alcock_2001,Afonso_2003}) and theoretical lower limits on the long-term survivability of hydrogenous objects (which rule out MACHOs with $M\lesssim 10^{-7} M_{\odot}$; \citealt{DJM92}), the Y04 wide binary constraints ruled out almost all possible MACHO masses. Only a small window of MACHO masses, $30 \lesssim M/M_{\odot} \lesssim 43$, was still considered viable. However, \citet{Quinn09} later argued that the Y04 constraints depend critically on the validity of the two widest binaries in the CG04 sample. After removing one spurious candidate binary from the CG04 sample, they repeated the analysis of Y04 and found the wide binary separation distribution consistent with a MACHO mass of $30 \lesssim M/M_{\odot} \lesssim 500$. 

Precise proper motions and parallaxes from Gaia DR2 \citep{gaia2016, gaia2018} have allowed the construction of larger and purer wide binary catalogs than were available previously. Distance measurements from Gaia parallaxes have also made it possible to measure the {\it physical} separation distribution; this is more easily interpretable than the angular separation distribution, whose physical scale depends on distance. \citet[][hereafter ER18]{El-Badry2018a} searched Gaia DR2 for high-confidence wide binaries within $200$\,pc of the Sun. Their search yielded $\sim$55,000 high-confidence wide binaries, including $\sim$3,500 in which at least one component is a white dwarf. A similar search was carried out by \citet{Pittordis_2019}, who used the projected orbital velocities of Gaia wide binaries to constrain the gravitational force law in the low acceleration regime. Both of these studies limited their search to projected separations of $s< 50,000$\,AU (0.25 pc), primarily because the contamination rate from chance alignments would have become non-negligible at wider separations given their selection criteria. Because they searched for nearby binaries without any kinematic selection, the majority of the binaries in these samples have disk-like kinematics and roughly solar metallicity \citep[see][]{EBR2019}.

Fitting the separation distribution with a broken power-law model, ER18 found that for MS/MS binaries, it was nearly consistent with a single power law of logarithmic slope $\gamma \approx -1.6$ but displayed evidence of a weak break at $\log(\rm s/AU) \approx 3.8$. They found the separation distributions for binaries containing a white dwarf to fall off more steeply at large separations and interpreted this steepening as evidence of non-adiabatic and/or asymmetric mass loss in the end stages of stellar evolution. The effects of disruption due to external perturbations are expected to become significant only at very wide separations ($s\gtrsim 20,000$\,AU; \citealt{Weinberg_1987}), so ER18 did not attempt to constrain the effects of external perturbations using their sample. 

In this paper, we (a) extend the ER18 analysis to wider separations and (b) compare the separation distributions of binaries between the Milky Way disk and halo. Constructing a large sample of halo binaries requires us to search to larger distances than ER18, but we show that it is still possible to obtain a high-purity binary sample at larger distances for halo stars, which have large proper motions and thus have fewer nearby neighbors in phase space. We kinematically select three pure subsamples with different average ages, using $v_{\perp,\rm tot}$, the total tangential velocity with respect to the Sun, as a proxy of stellar age. By investigating the three binary populations, we try to answer three basic questions: (1) Is there a slope change in the separation distribution of wide binaries at $s\approx 10,000$\,AU, as found tentatively by ER18? (2) If so, does its strength vary between stellar populations? (3) Can it provide a meaningful constraints or clues on formation or destruction mechanisms of binaries in different populations? 

Most previous works using wide binaries as dynamical tracers have implicitly assumed that the primordial separation distribution can be well-approximated as a simple power law at wide separations, and that it is the same for different populations. This is a serviceable assumption, and most previous works have found a power-law separation distribution to provide a good fit for wide binaries in the range of separations where external perturbations are expected to be subdominate \citep{Lepine_2007, Andrews_2017}, but there is little {\it a prior} motivation for it. Now that the present-day separation distribution can be constrained in more detail, we reexamine this assumption.
 
The remainder of this paper is organized as follows. In Section 2, we describe how to build an initial the wide binary candidate catalog that is extensive but not pure. Sections 3 then presents three pure subsamples with different stellar populations selected from the candidate catalog, i.e., the disk-like, the kinematically intermediate, and the halo-like binaries. In Section 4, we illustrate the observed separation distributions for the three subsamples. In order to infer the intrinsic separation distributions, we describe the method in Section 5, including the selection function of the observed sample, a  smoothly broken power law parameterization, and the likelihood for fitting the separation distribution. Section 6 lists the key results of the investigation. We discuss possible theoretical interpretations of our results in Section 7 and summarize in Section 8. 

Throughout the paper, we adopt the Solar motion as $(U_\odot,V_\odot,W_\odot)=(9.58, 10.52, 7.01)$\,\kms\ \citep{tian2015}, and the circular speed of the local standard of rest (LSR) as $v_0=238$\kms \citep{schonrich2012}.  $\alpha^{*}$ is used to denote the right ascension in the gnomonic projection coordinate system, for example,  $\mu_{\alpha^{*},i}\equiv \mu_{\alpha,i}\cos\delta_{i}$, and $\Delta\alpha^{*}\equiv\Delta\alpha\cos(\delta)$.

\section{The Initial Wide Binary Candidate Catalog}
\label{sect:data}
We first construct a catalog of wide binary candidates that is large but not pure. In Section 3, we describe how pure subsamples can then be selected from the initial candidate catalog.

We select wide binary candidates with a procedure  similar to that of ER18, whose catalog was restricted to pairs of stars within 200\,pc of the Sun and projected separations $s < 0.25$\,pc. They selected stars whose positions, proper motions, and parallaxes were consistent with being gravitationally bound. For the present study, we extend the search volume from 200\,pc to 4.0\,kpc and the maximum projected separation from $s = 0.25$\,pc to $s = 1.0$\,pc. We briefly summarize the selection procedure below.

\subsection{General criteria for the wide binary query}\label{sect:general}
In the first step, we search for wide binary candidates that satisfy the following criteria:
\begin{enumerate}
    \item \texttt{parallax} $>0.25$, and \texttt{parallax\_over\_error} $>20$ for the primary star. Possible companions are searched in a circle corresponding to a projected radius 1.0\,pc around each primary, within an angular separation ${\theta/}{{\rm arcsec}}\leq 206 \times{\varpi/}{{\rm mas}}$, where $\varpi$ is the parallax of the primary star. The secondary companions are required to meet a lower threshold for parallax error,  \texttt{parallax\_over\_error} $>2$ (lower than the cut in ER18).

    \item Consistent distances for the primary and secondary. We require $\Delta d\leq3\sigma_{\Delta d} + 2s$, where $\Delta d=\left|1/\varpi_1-1/\varpi_2\right|$ is the difference in distance between the two stars,  $\sigma_{\Delta d}=(\sigma_{\varpi,1}^{2}/\varpi_{1}^{4}+\sigma_{\varpi,2}^{2}/\varpi_{1}^{4})^{1/2}$ is its uncertainty, and $\varpi_i$ and $\sigma_{\varpi, i}$ represent the parallax of a primary or secondary and reported uncertainty. The $+2s$ term prevents us from missing the nearest and widest binaries, for which the orbital separation can be an non-negligible fraction of the distance.

    \item Small proper motion differences between the two stars, consistent with a bound Keplerian orbit. We require $\Delta\mu\leq\Delta\mu_{{\rm orbit}}+3.0\sigma_{\Delta\mu}$, where $\Delta \mu = \left[(\mu_{\alpha,1}^{*} - \mu_{\alpha,2}^{*})^2 + (\mu_{\delta, 1} - \mu_{\delta, 2})^2\right]^{1/2}$, which is the total scalar difference in proper motion between the two stars and $\sigma_{\Delta\mu}=\frac{1}{\Delta\mu}\left[\left(\sigma_{\mu_{\alpha,1}^{*}}^{2}+\sigma_{\mu_{\alpha,2}^{*}}^{2}\right)\Delta\mu_{\alpha}^{2}+\left(\sigma_{\mu_{\delta,1}}^{2}+\sigma_{\mu_{\delta,2}}^{2}\right)\Delta\mu_{\delta}^{2}\right]^{1/2}$ is its uncertainty. $\frac{\Delta \mu_{{\rm orbit}}}{{\rm mas\,yr}^{-1}} = 0.44\left(\frac{\varpi}{{\rm mas}}\right)^{3/2}\left(\frac{\theta}{{\rm arcsec}}\right)^{-1/2}$ represents the maximum proper motion difference permissible for a circular orbit of total mass $5\,M_{\odot}$. We also require the proper motions to be reasonably precise, with $\sigma_{\Delta\mu}\leq 1.5\,{\rm mas\,yr^{-1}}$. 
        
\end{enumerate}

\subsection{Quality cuts}\label{sect:quality}
We apply additional quality cuts on the astrometry and photometry of both components of candidate binaries: 
\begin{enumerate}
    \item $\sqrt{\chi^2/(\nu'-5)}<1.2\times \max(1, \exp(-0.2(G-19.5))$, to make sure that both members of a candidate binary pair have an acceptable astrometric solution \citep{Lindegren2018}. $\chi^2$ and $\nu'$ are respectively referred to as \texttt{astrometric\_chi2\_al} and \texttt{astrometric\_n\_good\_obs\_al} in the {\it Gaia} archive.
    \item  $1.0+0.015({\rm G_{BP}-G_{RP}})^{2} <$ \texttt{phot\_bp\_rp\_excess\_factor} $ < 1.3+0.06({\rm G_{BP}-G_{RP}})^{2}$, to ensure that both stars have photometry that is relatively uncontaminated by nearby sources \citep{Evans2018}.
    \item \texttt{phot\_g\_mean\_flux\_over\_error}$>$50 for both member stars, \texttt{phot\_rp\_mean\_flux\_over\_error}$>$20 ($>$10) for the primary (secondary) star, and \texttt{phot\_bp\_mean\_flux\_over\_error}$>$20 ($>$10) for the primary (secondary) star, to remove pairs with low-SNR photometry.
\end{enumerate}
These selection criteria yield an initial sample of 16,973,885 wide binary candidates.

\subsection{Removing clusters, moving groups, and pairs in high density regions}
Along with wide binaries, the criteria specified in Sections \ref{sect:general} naturally select stars in bound clusters and moving groups. We remove these using a similar approach to that adopted by ER18.

For each candidate binary, we define nearby neighbor binaries as those that are within 1 degree on the sky, $\pm$3\masyr\ in both proper motion coordinates, and $\pm$5\,pc in $1/\varpi$. According to this definition, we count the number ($N$) of nearby neighbors in position-parallax-proper motion space, and remove candidate pairs that have $N>10$ nearby neighbors. 16,166,274 ($>95\%$) pairs are removed in this step. 

Because our binary candidate catalog is larger and contains fainter stars than the one constructed in ER18, the practical effect of this cut is different from the one applied in ER18. There, most of the candidate pairs removed were members of bound clusters. The cut here also removes clusters, but a large majority of the removed candidates are simply pairs in regions of high stellar density. As we discuss below, the contamination rate is high for such pairs anyway, so removing them is an acceptable concession. 

This leaves 807,611 pairs as wide binary candidates.

\subsection{Properties of the initial candidates}\label{sec:summary}
\begin{figure}[!t]
\centering
\includegraphics[width=0.2565\textwidth, trim=0.0cm 0.0cm 0.0cm 0.0cm, clip]{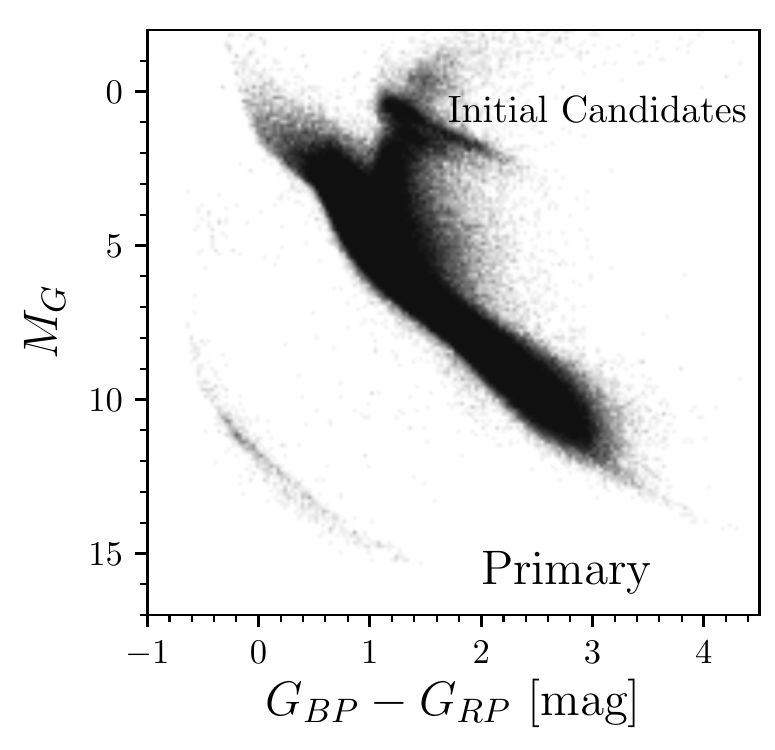}
\includegraphics[width=0.2108\textwidth, trim=1.4cm 0.0cm 0.0cm 0.0cm, clip]{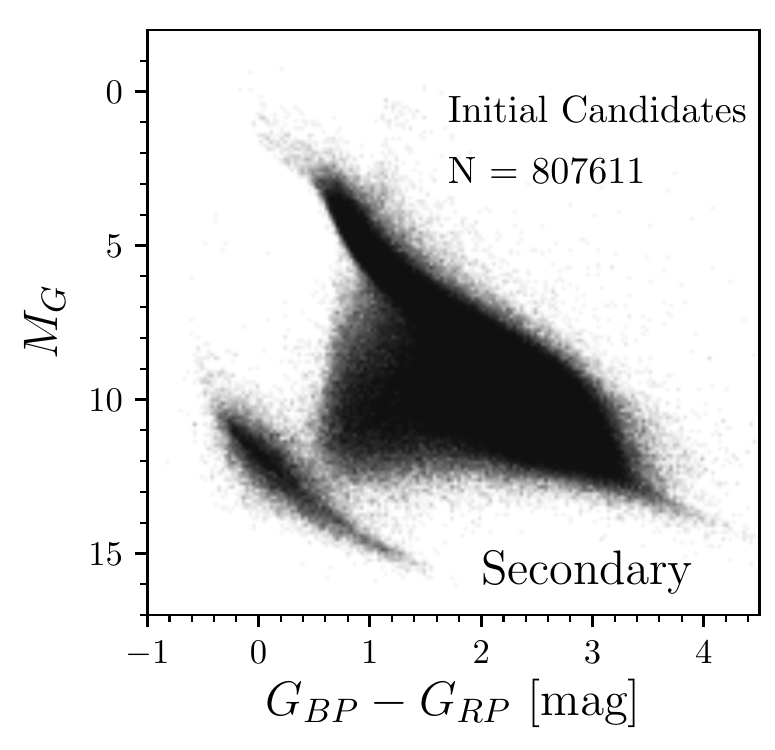}
\includegraphics[width=0.2565\textwidth, trim=0.0cm 0.0cm 0.0cm 0.0cm, clip]{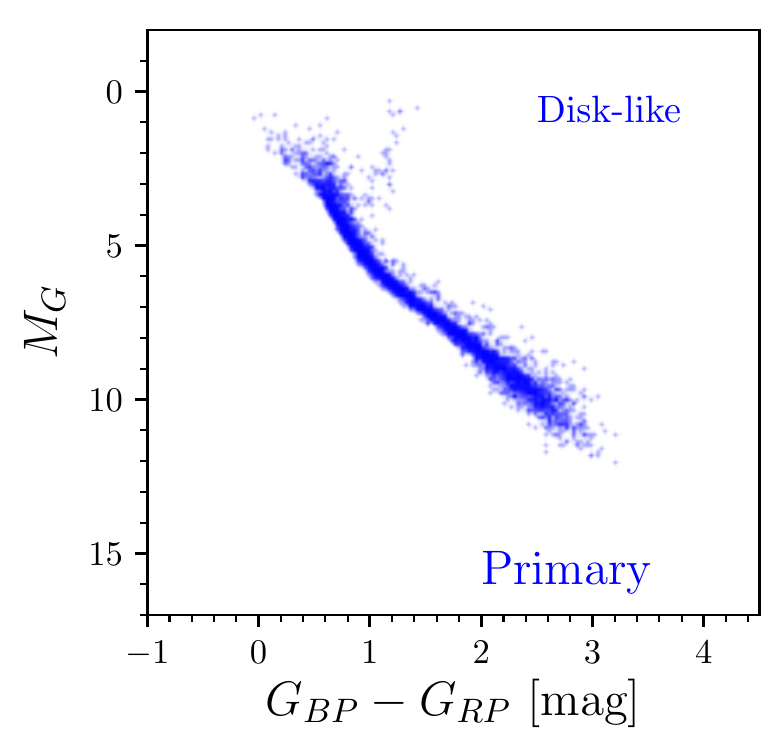}
\includegraphics[width=0.2108\textwidth, trim=1.4cm 0.0cm 0.0cm 0.0cm, clip]{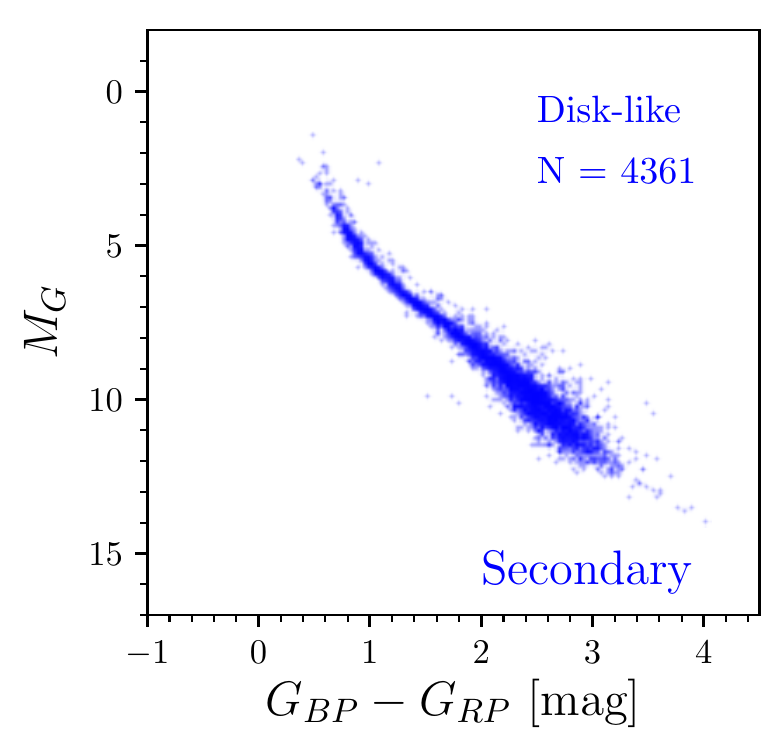}
\includegraphics[width=0.2565\textwidth, trim=0.0cm 0.0cm 0.0cm 0.0cm, clip]{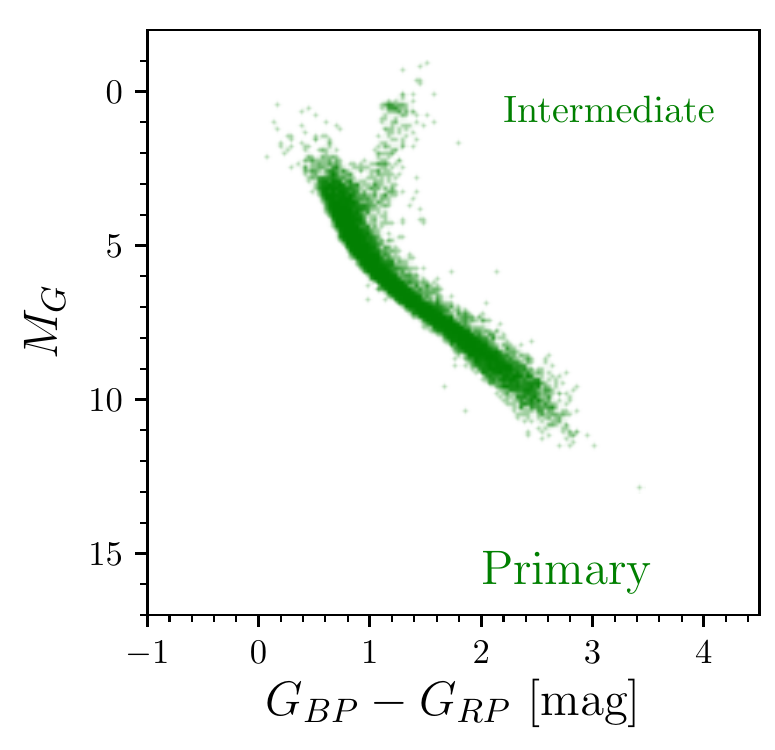}
\includegraphics[width=0.2108\textwidth, trim=1.4cm 0.0cm 0.0cm 0.0cm, clip]{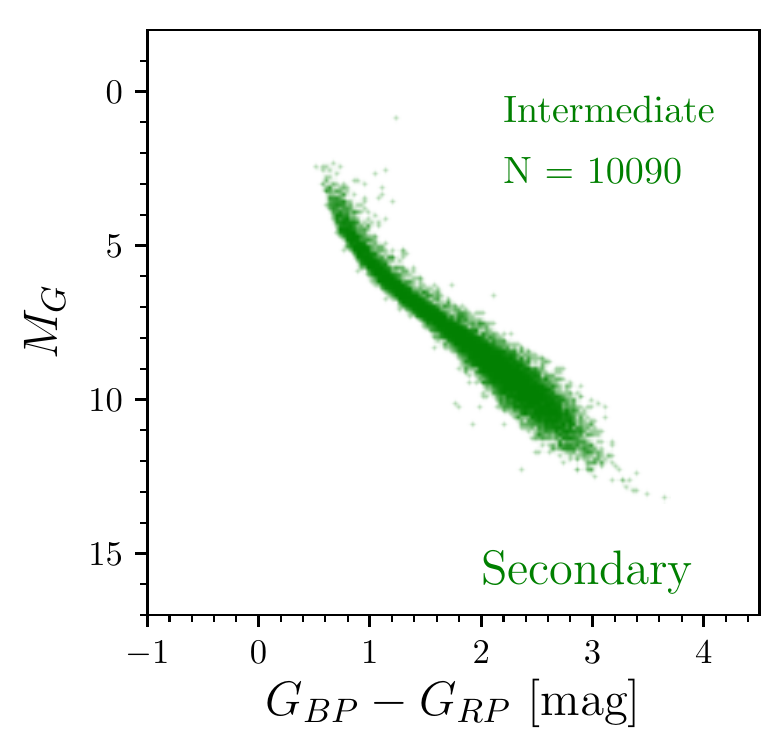}
\includegraphics[width=0.2565\textwidth, trim=0.0cm 0.0cm 0.0cm 0.0cm, clip]{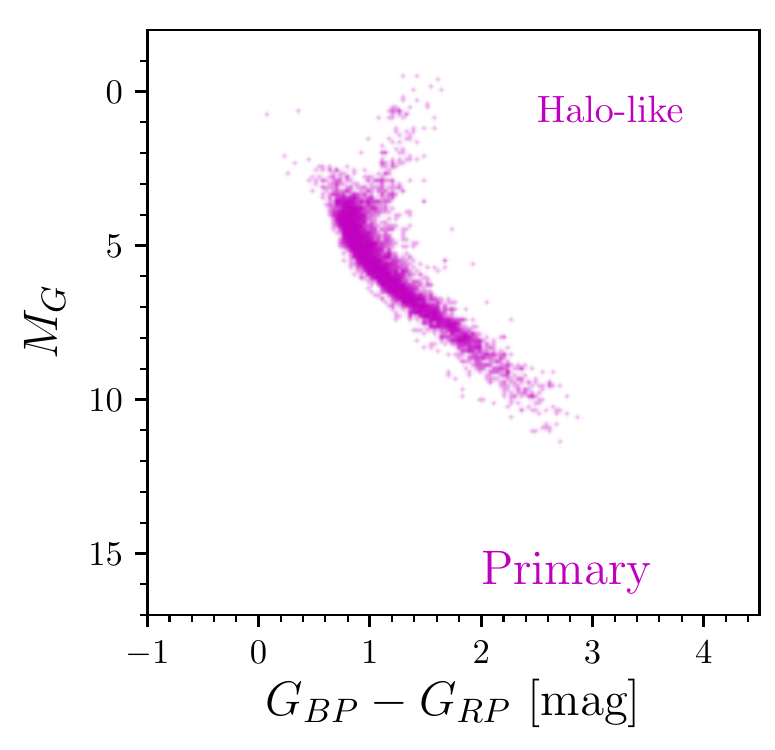}
\includegraphics[width=0.2108\textwidth, trim=1.4cm 0.0cm 0.0cm 0.0cm, clip]{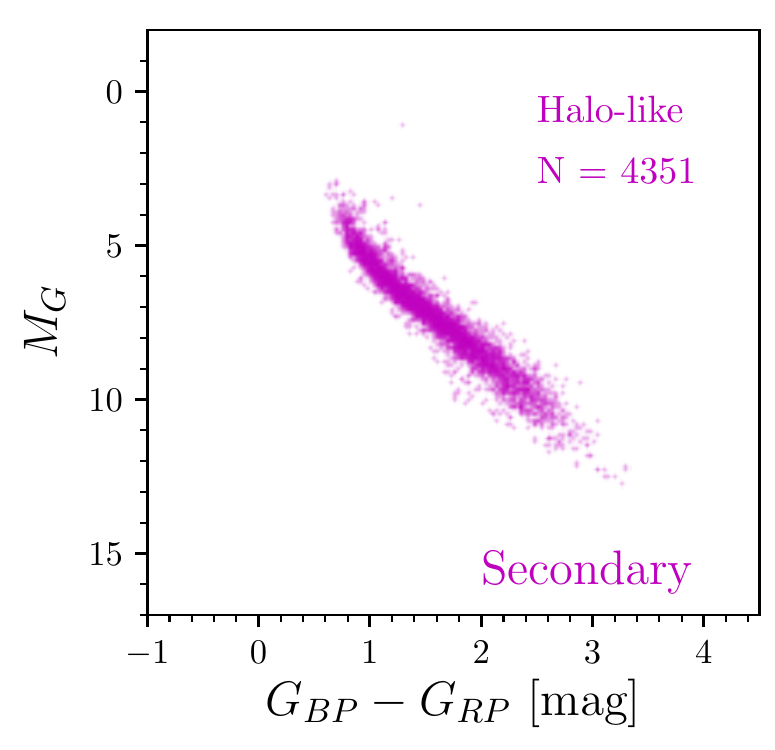}
\caption{Color-magnitude diagrams (CMDs) for the observed binaries.  All the initial binary candidates (807,611 pairs) are displayed with the black dots, while the blue (4361 pairs), magenta (4351 pairs), and the green (10,090 pairs) dots show  disk-like, intermediate, and halo-like MS-MS wide binaries, respectively. Here the ``primary" and ``secondary" label the brighter and fainter members, respectively.}\label{fig:CMD_sep}  
\end{figure}

Figure \ref{fig:CMD_sep} (top panel) shows a color-magnitude diagram (CMD) for all  807,611 (pairs) of wide binary candidates. The (primary) stars with $\rm M_G < 2.75(G_{BP}-G_{RP})+5.75$ are designated as ``main sequence'' stars (a few giants are not excluded), and the objects with $\rm M_G > 3.25(G_{BP}-G_{RP})+9.63$ are classified as likely white dwarfs (WDs). Here ${\rm M_{G}=G+5\log\left(\varpi/mas\right)-10}$. The main sequence and red clump are visibly smeared out, primarily because the photometry is not corrected for extinction. Deviations from the expected CMD morphology are more pronounced for secondaries, (top right panel in Figure \ref{fig:CMD_sep}) because the cuts on astrometric and photometric precision (Section \ref{sect:quality}) are looser for secondaries. A large cloud of objects is visible between the white dwarfs and main sequence for secondaries. These are primarily sources with large parallax errors; they are removed from the sample once more stringent quality cuts are applied (See lower panels of Figure 1). 
A secondary sequence is apparent above the main sequence for both primaries and secondaries, indicating that the catalog contains hierarchical triples with components that are spatially unresolved close binaries. 

\begin{figure}[!t]
\centering
\includegraphics[width=0.45\textwidth, trim=0.0cm 0.0cm 0.0cm 0.0cm, clip]{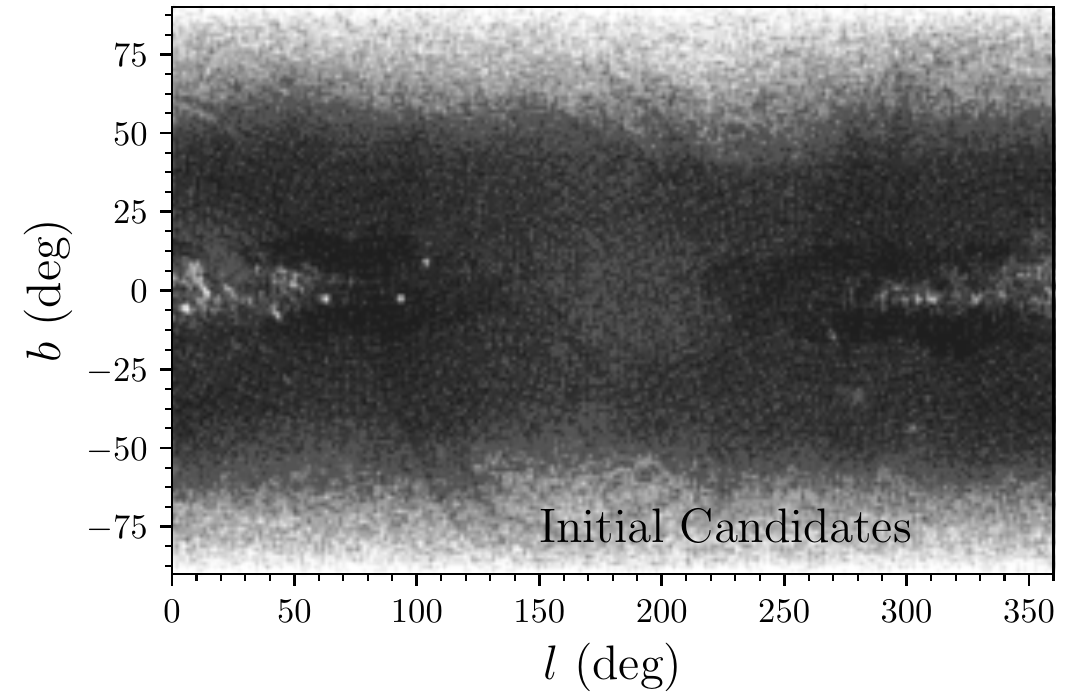}
\includegraphics[width=0.45\textwidth, trim=0.0cm 0.0cm 0.0cm 0.0cm, clip]{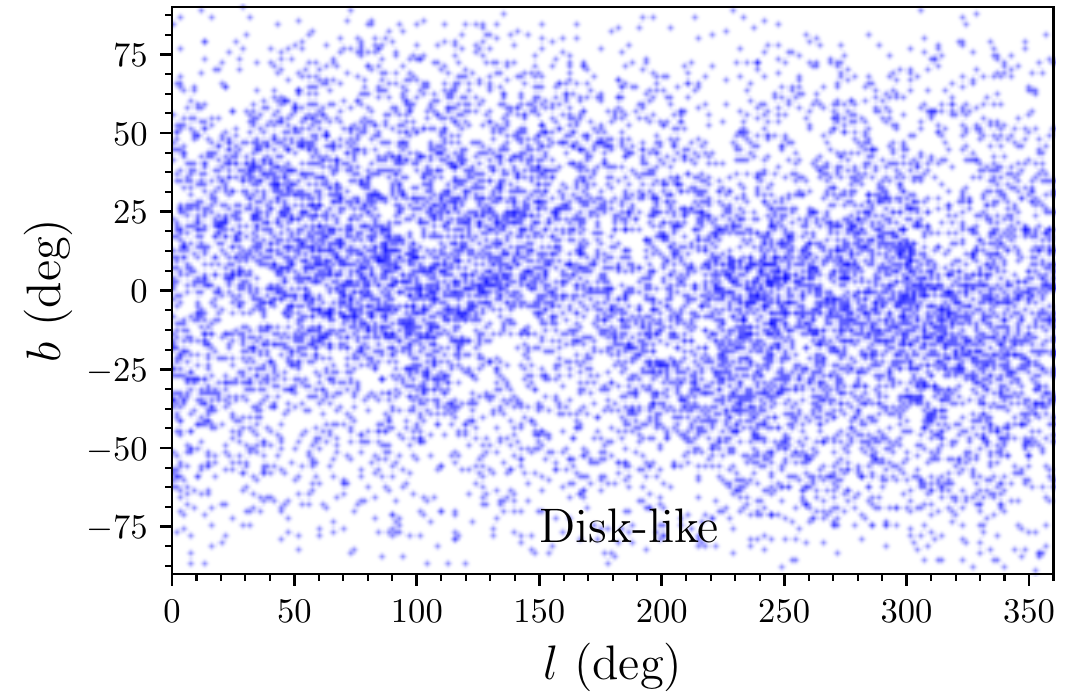}
\includegraphics[width=0.45\textwidth, trim=0.0cm 0.0cm 0.0cm 0.0cm, clip]{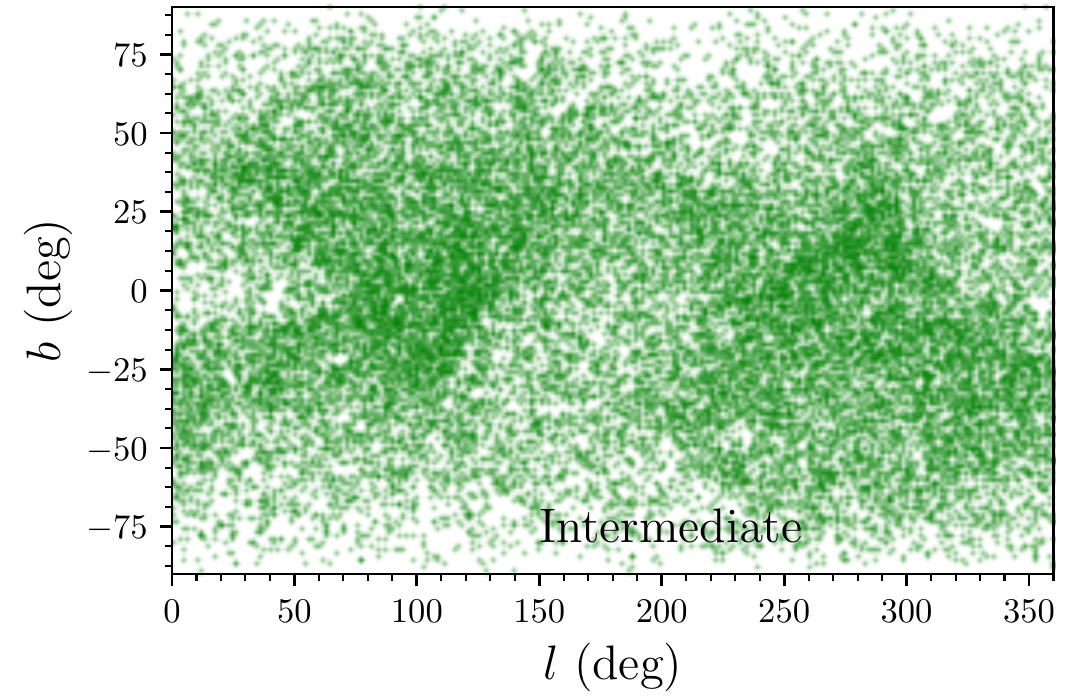}
\includegraphics[width=0.45\textwidth, trim=0.0cm 0.0cm 0.0cm 0.0cm, clip]{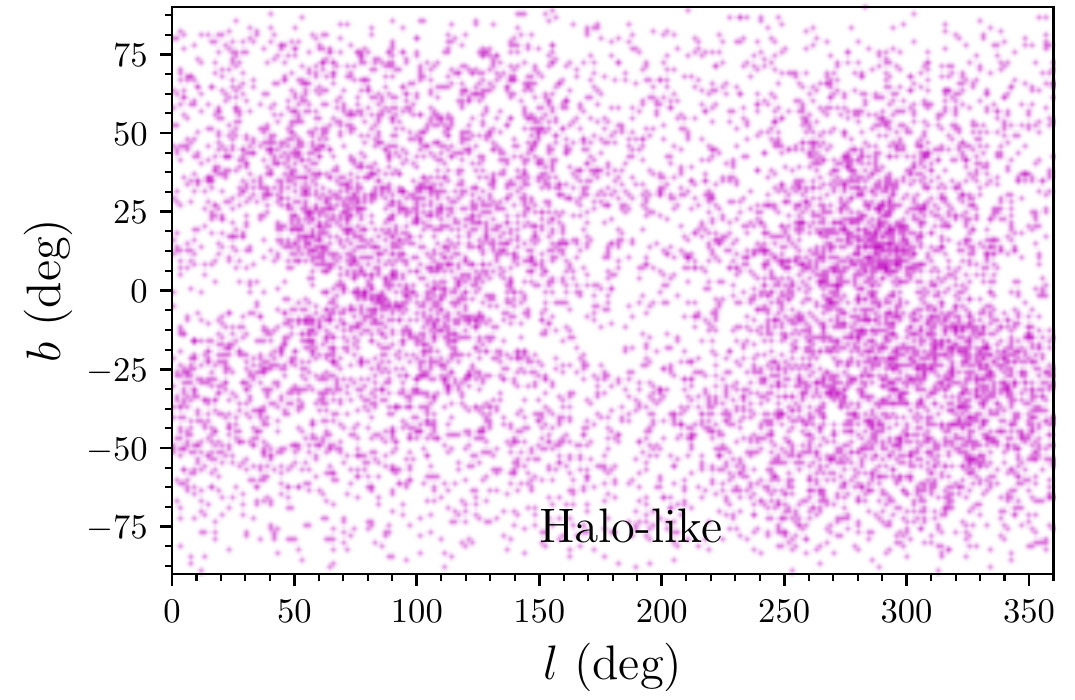}
\caption{Distribution of the observed binary samples in the $b$ vs. $l$ plane, including 807,611 initial wide binary candidates (black), 4361 high-confidence disk-like binaries (blue), 10,090 high-confidence intermediate binaries (green), and 4351 high-confidence halo-like binaries (magenta). Imprints of the {\it Gaia} scanning law can be seen in all samples. }\label{fig:lb_distr}
\end{figure}

Figure \ref{fig:lb_distr} (top panel) presents the sky distribution of all binary candidates in the $b$ vs. $l$ plane, where $b$ and $l$ represent galactic latitude and longitude. Clear imprints of the {\it Gaia} scanning law are apparent, as there are more sources with well-constrained astrometry in regions of the sky that were visited more often. An higher density of candidates is also apparent near the galactic plane; this was even more pronounced prior to the removal of sources with many phase-space neighbors. A few blank patches at $b\sim0\degr$ are very dense regions where our queries timed out and no candidates could be identified.

\begin{figure*}[!t]
\centering
\includegraphics[width=0.28\textwidth, trim=0.cm 0.0cm 0.0cm 0.0cm, clip]{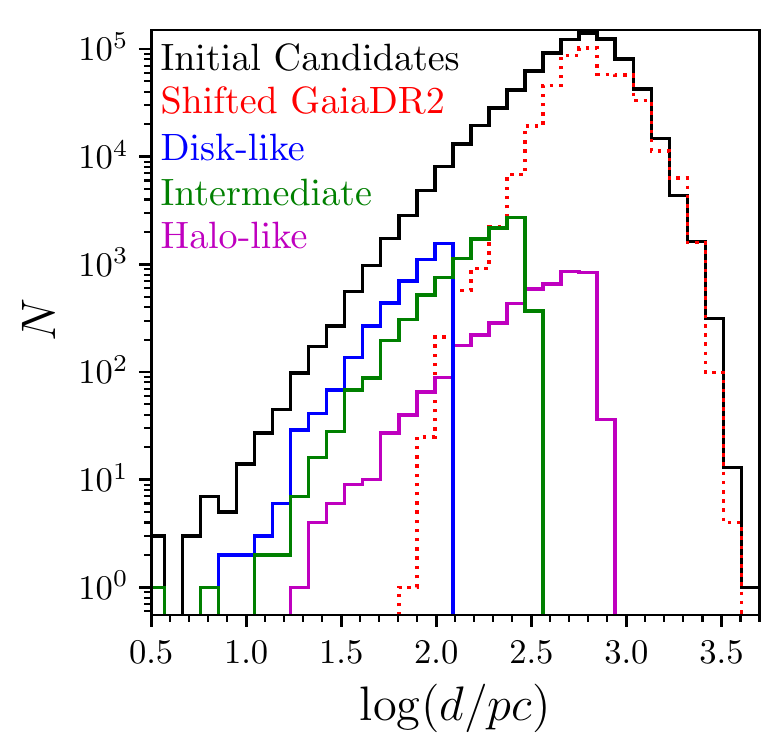}
\includegraphics[width=0.2275\textwidth, trim=1.43cm 0.0cm 0.0cm 0.0cm, clip]{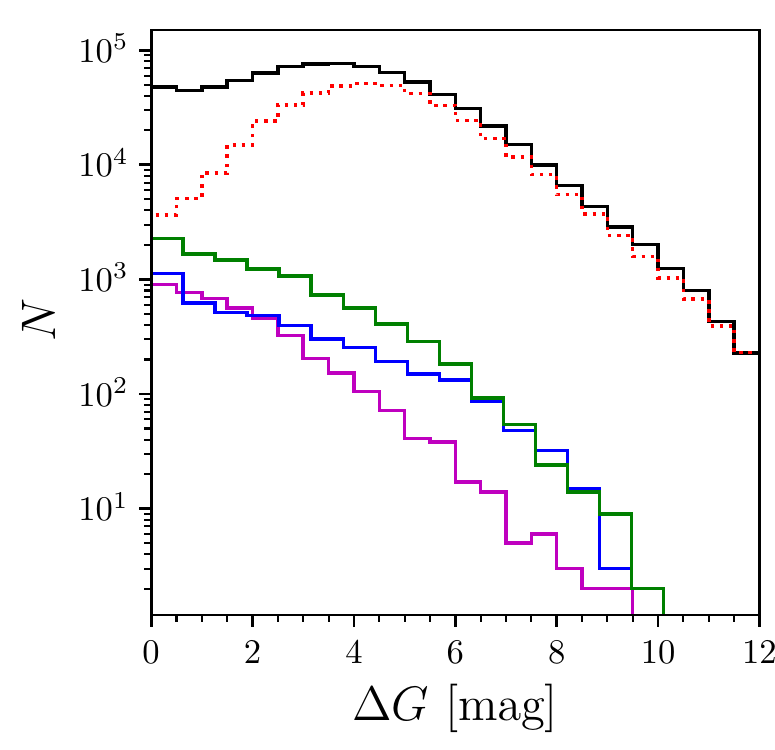}
\includegraphics[width=0.2275\textwidth, trim=1.43cm 0.0cm 0.0cm 0.0cm, clip]{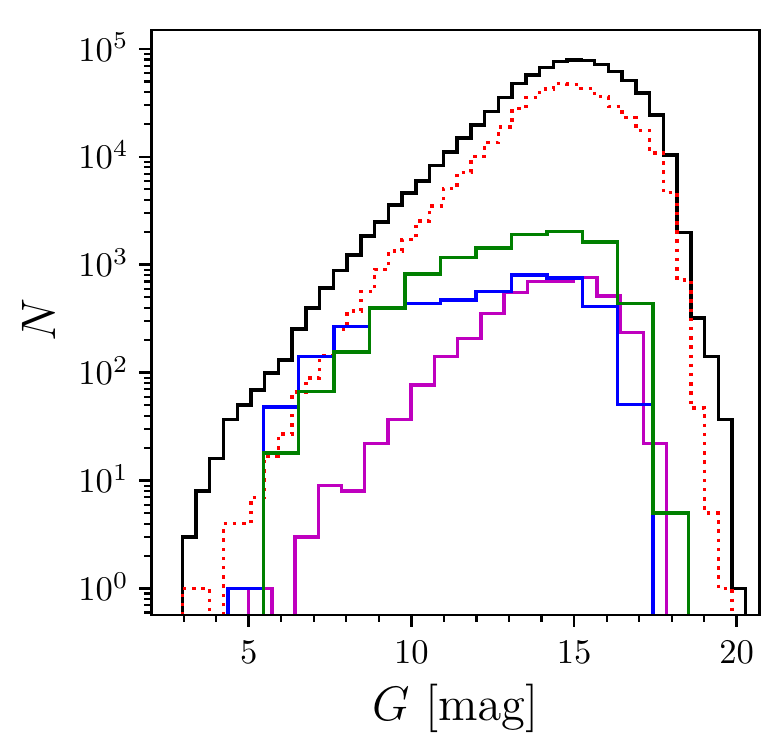}
\includegraphics[width=0.2275\textwidth, trim=1.43cm 0.0cm 0.0cm 0.0cm, clip]{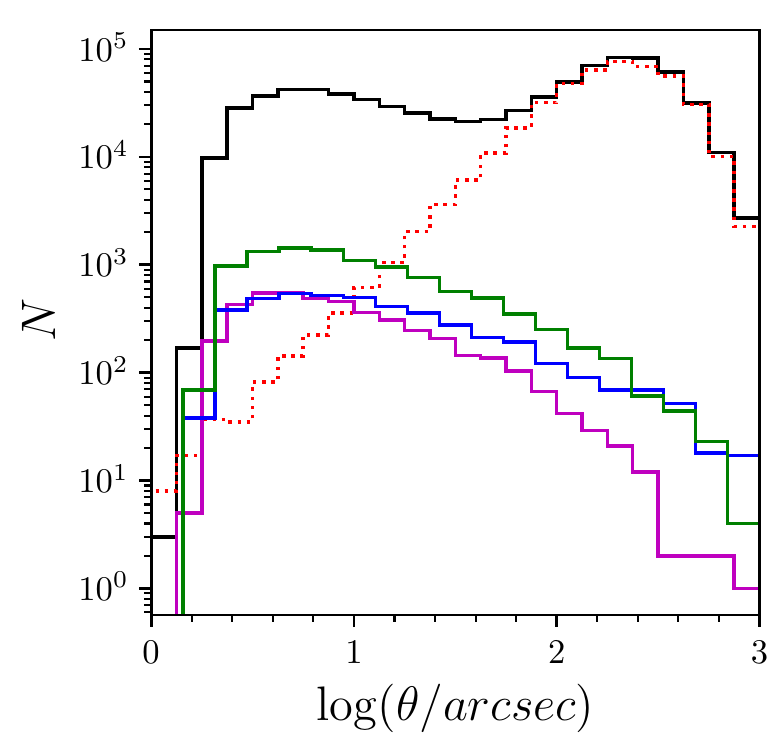}
\caption{Comparison of various wide binaries subsets and chance alignments. From left to right, the sub-panels show  distributions of distances ($\log(d)$, first column), magnitude difference ($\Delta \rm G=|G_1-G_2|$, second column), magnitude ($G$, third column), and angular separation ($\log\theta$, fourth column), for the whole candidate catalog (black), the clean disk-like subset (blue), the intermediate clean subset (green), and the halo-like clean subset (magenta). The red dotted curves display distributions of the chance alignments estimated from a shifted Gaia DR2 catalog. This catalog is produced by shifting all the objects of Gaia DR2 by 1\degr\ in both the right ascension and declination directions. Matching the original Gaia DR2 with the shifted catalog, the matched pairs are regarded as chance alignments if they pass our selection criteria for binaries. 
}\label{fig:hist_distrs}
\end{figure*}

\begin{figure}[!t]
\centering
\includegraphics[width=0.45\textwidth, trim=0.0cm 0.0cm 0.0cm 0.0cm, clip]{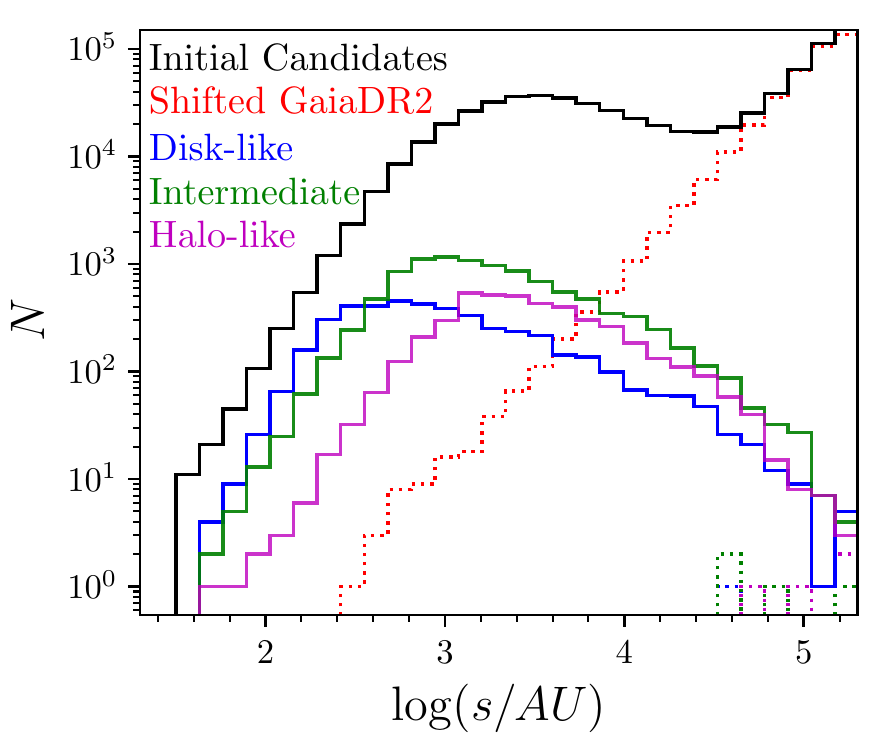}
\caption{Distributions of projected separation ($\log(s)$) for the initial candidate sample, the three clean subsets, and the chance alignments selected from the shifted Gaia DR2 catalog. The color codes are same as those in Figure \ref{fig:hist_distrs}. As an estimate of the contamination rate in the three clean subsets, the blue, green and magenta dotted curves display the distributions of the chance alignments selected from the shifted catalog with the disk-like, intermediate and halo-like criteria, respectively. 
}\label{fig:hist_distr_sAU}
\end{figure}

Figure \ref{fig:hist_distrs} illustrates the distributions of distance ($\log(d)$, first column), magnitude difference ($\Delta \rm G=|G_1-G_2|$, second column), apparent magnitude ($\rm G$, third column), and angular separation ($\log\theta$, fourth column) for all wide binary candidates (black). The distribution of {\it physical} separation is displayed separately in Figure \ref{fig:hist_distr_sAU}. For the initial candidate sample, the distributions are dominated by contaminants (chance-alignments) at large separations ($\theta\sim4.2$\,arcmin and $\rm s > 0.05$\,pc). This can be seen in the distributions of $\log\theta$ and $\log(s)$, where the large-separation peak is due entirely to chance alignments. We discuss chance-alignment further below.

\subsection{Contamination rates of the initial sample}

We use three methods to quantify the contamination rates of the candidate binaries as a function of separation: (1) A ``shifted Gaia DR2 catalog'', where the coordinates ($\hat{\rm \alpha}$, $\hat{\rm \delta}$) for each object in this catalog have been constructed by shifting 1\degr\ from its original location (${\rm \alpha}$, ${\rm \delta}$), i.e., ($\hat{\rm \alpha}$, $\hat{\rm \delta}$) = (${\rm \alpha}$ + $\Delta\alpha^{*}/\cos(\delta)$, ${\rm \delta}$ + $\Delta\delta$), with $\Delta\alpha^{*}=\Delta\delta=1.0$\degr. We then repeat the binary candidate identification procedure, now identifying pairs that pass the binary cuts when the coordinates of the ``primary'' are shifted relative to candidate secondaries. This procedure removes genuine binaries, but preserves chance alignment statistics; see \citet{Lepine_2007} for further discussion of this method of estimating the chance alignment rate. (2) The mock Gaia DR2 catalog produced by \citet{Rybizki2018}, which is based on the Besan\c con stellar population synthesis model \citep{Robin2003} and is populated with the Galactic distribution function using Galaxia \citep{Sharma2011}, assuming a similar selection function and uncertainty model to Gaia DR2. (3) Comparison of radial velocities (RVs) for the subset of bright wide binaries where both stars have RVs included in {\it Gaia} DR2. 

The red dotted curves in Figure \ref{fig:hist_distrs} present the distributions of the random alignment pairs chosen from the shifted Gaia DR2 catalog in the whole sky. We matched the original Gaia DR2 with the shifted catalog, and regard the matched pairs as chance alignments if they pass the selection criteria for binaries. This approach is not appropriate for the widest and most nearby binaries, which can have projected separations of $1^\circ$ or larger, and so continue to appear as binaries even when the coordinates
of one component are artificially shifted by $1^\circ$ in each
direction. Therefore, we remove the chance alignments with d $<$ 60 pc and $\theta >$ 0.5\degr\ from the shifted catalog. The red dotted curves suggest that the distributions for the initial candidates are indeed almost fully dominated by contaminants at large angular and projected separations, as displayed in the last two sub-panels of Figure \ref{fig:hist_distrs}. On the other had, chance alignments are subdominant at $\log(s) \lesssim 4.5$ (Figure~\ref{fig:hist_distrs}). The distribution of chance alignments can be well described as $N \sim  s\,{\rm d}s$, since  the area within which chance alignments can be found scales as $2\pi s\,{\rm d}s$.  

We selected in a random region of $\sim 4000\ \rm deg^2$ from the mock Gaia DR2 catalog \citep{Rybizki2018}. The mock catalog does not contain any true binaries, so any pairs that pass our selection criteria in the mock catalog must be chance alignments. We only query a fraction of the sky in the mock catalog because querying it is computationally expensive and many of the queries time out. In estimating the true contamination rate, we scale up the rate predicted by the mock catalog at all separations, assuming the contamination rate approaches $\sim $100\% at 1~pc separations.

\begin{figure}[!t]
\centering
\includegraphics[width=0.4\textwidth, trim=0.0cm 0.0cm 0.0cm 0.0cm, clip]{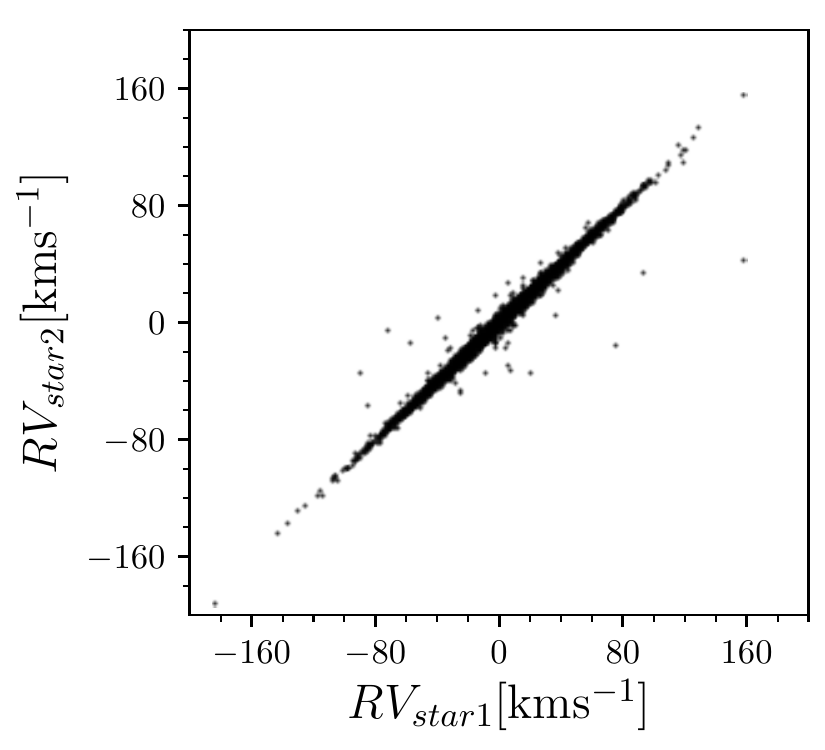}
\includegraphics[width=0.4\textwidth, trim=0.0cm 0.0cm 0.0cm 0.0cm, clip]{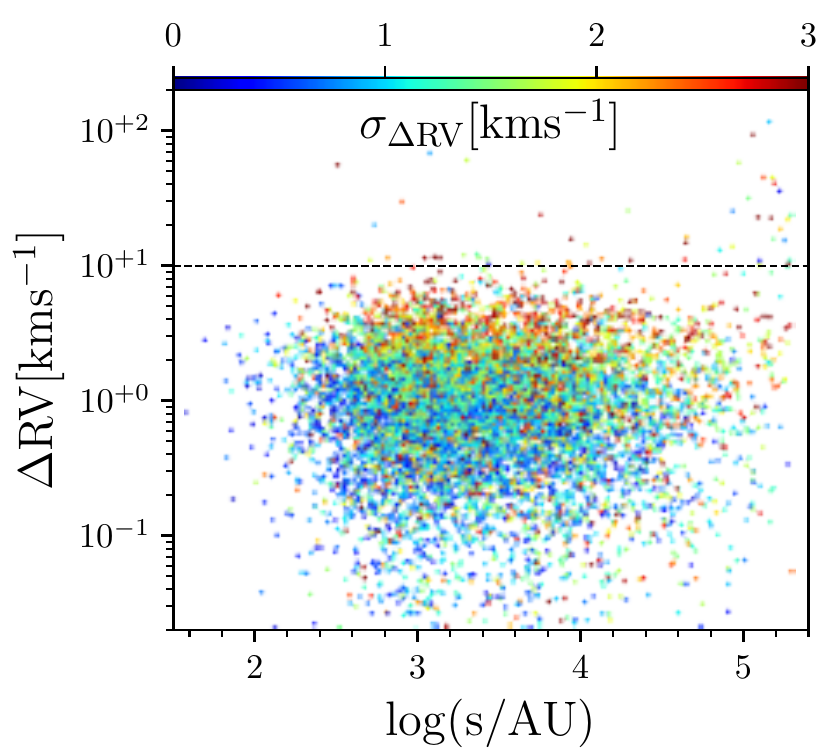}
\caption{Radial velocity 1-to-1 comparison (top) and $\rm \Delta RV$ vs. $\rm \log(s/AU)$ scatter (bottom) for 8220 binary candidates. The reliable $\rm RVs$ with uncertainties $\rm \sigma_{RVs}<3$\,kms\ are obtained from Gaia DR2. 28 pairs have $\rm \Delta RV>10$\,\kms, 140 pairs have $\rm \Delta RV$/$\rm \sigma_{\Delta RV}>5$, and only 20 pairs meet both conditions. Most of them are at large separations ($\rm s>10^{4.9}$\,AU). The black dashed line marks the location of $\rm\Delta RV=10$\,\kms.}\label{fig:rv_drv}
\end{figure}

In defining candidate wide binaries in this study, we  used the two dimensional velocities (i.e., 2-D tangential velocities from proper motions), positions and distances. The incomplete kinematic information induces substantial contamination, particularly at large separations. Gaia DR2 provides about 7.2 million stars with measured RVs. Among them, 8220 wide binary candidates have reliable (uncertainty $\rm \sigma_{RV}<3$\,\kms) RVs for both binary components. These RVs provide the third dimension of velocity and can be effectively used to check whether candidate pairs are true binaries or chance alignments. Significant RV differences ($\Delta \rm RV$) between member stars imply that star pairs are probably chance alignments. We regard those pairs with $\Delta \rm RV/\sigma_{RV}>5$ and $\Delta \rm RV>10$ \kms\  as likely contaminants. Figure \ref{fig:rv_drv} displays the 1-to-1 scatter distribution (the top panel) of the two component RVs, and the $\rm \Delta RVs$ distribution (the bottom panel) as a function of projected separation. Only 20 pairs meet the criteria of chance alignments, i.e., $\Delta \rm RV/\sigma_{RV}>5$ and $\Delta \rm RV>10$ \kms. Most binaries with substantially large $\Delta \rm RV$ have larger-than-average RV uncertainties. This implies that the contamination rate of the candidate binary catalog is of order 0.24\%, with most contaminants at $\log(s/{\rm AU})>5$, {\it for binaries that are bright enough for both stars to have measured RVs ($G\lesssim 13$)}. However, we stress that the true contamination rate for the whole catalog of binary candidates is much higher, as can be seen in Figure~\ref{fig:hist_distrs}. The RVs provide an estimate of the chance-alignment rate for binaries where both components are bright. 

Table \ref{tab:contami_rate} summarizes the contamination rates in the different projected separation bins, estimated with the above three methods.

\begin{table*}
\begin{threeparttable}
\caption{Contamination rates of the wide binary candidates and cleaned subsamples estimated by different methods}.\label{tab:contami_rate}
\centering
\begin{tabular}{c|c|c|c|c|c|c|c|c|c|c|c|cc}
\hline
\hline
Separations&\multicolumn{12}{c}{Contamination Rates}\\
\hline
log(s/AU)&\multicolumn{4}{c|}{Mock\footnotemark} &\multicolumn{4}{c|}{Gaia RVs\footnotemark}   &\multicolumn{4}{c}{Shifted Gaia DR2\footnotemark}      \\
\hline
&All&Disk&Intermediate&Halo&All&Disk&Intermediate&Halo&All&Disk&Intermediate&Halo\\
\hline
(0.00, 3.40]&0\%&0\%&0\%&0\%&0\%&0\%&0\%&0\%&$<$0.3\%&0\%&0\%&0\%\\
(3.40, 4.12]&$<$2\%&0\%&0\%&0\%&$<$0.1\%&0\%&0\%&0\%&$<$2\%&0\%&0\%&0\%\\
(4.12, 4.39]&$\sim$10\%&0\%&0\%&0\%&$<$0.1\%&0\%&0\%&0\%&$\sim$10\%&0\%&0\%&0\%\\
(4.39, 4.65]&$\sim$30\%&0\%&0\%&0\%&$<$0.2\%&0\%&0\%&0\%&$\sim$30\%&0\%&0\%&0\%\\
(4.65, 4.91]&$\sim$80\%&0\%&0\%&0\%&$\sim$0.4\%&0\%&0\%&0\%&$\sim$80\%&$\sim$1\%&$\sim$0.1\%&$\sim$1\%\\
(4.91, 5.20]&$>$90\%&0\%&0\%&0\%&$\sim$1.5\%&0\%&0\%&0\%&$>$90\%&0\%&$\sim$1\%&$\sim$5\%\\
(5.20, 5.31]&$>$95\%&0\%&0\%&0\%&$\sim$11\%&0\%&0\%&0\%&$>$95\%&0\%&$\sim$10\%&$\sim$15\%\\
\hline
\hline
\end{tabular}
 \begin{tablenotes}
  \item [a] Using the mock catalog from \citet{Rybizki2018} to estimate the contamination rate, assuming the contamination for the initial candidates reaches 100\% at the widest separations.
  \item [b] Using the reliable ($\rm\sigma_{RV}<3$ \kms) Gaia radial velocities (RVs) to estimate the contamination rate of the the brightest wide binaries, where both components have $G \lesssim 13$. Pairs with $\Delta \rm RV/\sigma_{RV}>5$ and $\Delta \rm RV>10$ \kms\ are regarded as chance alignments. 
  \item [c] Using the shifted Gaia DR2 to estimate the contamination rate. Each object is shifted by 1\degr\ in each coordinate i.e., ($\hat{\rm \alpha}$, $\hat{\rm \delta}$) = (${\rm \alpha}$ + 1.0\degr$/\cos(\delta)$, ${\rm \delta}$ + 1.0\degr), where ($\hat{\rm \alpha}$, $\hat{\rm \delta}$) and (${\rm \alpha}$, ${\rm \delta}$) are the shifted and original coordinates, respectively.
  \end{tablenotes}
 \end{threeparttable}
\end{table*}
    
\section{Three Pure Binary Samples with Different Kinematic Ages}

As low contamination is crucial for our subsequent analysis of the separation distributions, we proceed to define subsamples that we can assume to be nearly pure.
Specifically, we kinematically select three pure subsamples, consisting of young disk-like, intermediate, and old halo-like binaries, with different average ages. Since the phase-space density of contaminants varies between the samples (with more contaminants in the low-velocity samples), we use different quality cuts for the three samples. 

Using the Gaia DR2 proper motion and parallax, we calculate the total tangential velocity with respect to the Sun for each candidate binary:
\begin{equation}\label{eq:disk} 
v_{\perp,\rm tot}  \equiv 4.74 \rm km/s \times (\mu_{\rm tot}\times \rm yr)/\varpi.
\end{equation}
Here $\varpi$ and $\mu_{\rm tot}=\sqrt{\mu_{\alpha^*}^2 + \mu_{\delta}^2}$ are the parallax and total proper motion of a binary, respectively.

Using $v_{\perp,\rm tot}$ as a proxy of age, we select three pure subsamples with different average ages. We determine what cuts are needed to obtain a pure subsample in each population using the shifted catalog the estimated the chance alignment rate given any sub-selection.

The space density of halo stars in the solar neighborhood is much lower than that of disk stars, so it is necessary to search to larger distances to obtain a large sample of halo binaries. Fortunately, the contamination rate at fixed distance is also much lower for halo stars, because they have fewer phase-space neighbors. We thus use different distance limits for the three populations and self-consistently estimate the contamination rate for each population given these limits. We choose disk-like binaries with
  \begin{enumerate}
    \item d $<$ 120\, pc, 
    \item $v_{\perp,\rm tot} < 40$\, \kms.
\end{enumerate} 

To select the intermediate-age binary subsample, we use the conditions:
  \begin{enumerate}
    \item d $<$ 300\, pc, 
    \item $40< v_{\perp,\rm tot} < 85$\, \kms.
\end{enumerate} 

We choose old halo-like binaries with the following criteria,
  \begin{enumerate}
    \item d $<$ 700\,pc, 
    \item $v_{\perp,\rm tot} > 85$\, \kms.
\end{enumerate} 

To get pure binary samples, we further impose the following cuts on all of the above three selections:
  \begin{enumerate}
   \item $\Delta\mu\leq\Delta\mu_{{\rm orbit}}+1.0\sigma_{\Delta\mu}$ and $\sigma_{\Delta\mu}\leq 0.12\,{\rm mas\,yr^{-1}}$. This more stringent cut on proper motion uncertainty reduces the contamination from chance alignments with larger uncertainty. 
    \item  The number of nearby neighbors $N<2$, to strictly remove contaminants at wide separation from moving groups or star clusters.
     \item We exclude binaries containing a WD from all three subsamples to remove the effect from internal orbital evolution.
\end{enumerate}

The above criteria can effectively suppress the three kinds of contamination rates close to zero for the three subsamples at each (except the largest) separation bin (see Table \ref{tab:contami_rate}).
Finally, we get 4361 disk-like, 10,090 intermediate and 4351 halo-like genuine wide binaries. Figure \ref{fig:hist_distr_sAU} compares the separation distributions of binaries and chance-alignments in the three subsamples. With these more aggressive cuts on distance and astrometric SNR, chance alignments are subdominant out to separations of 1,pc. The separation distributions of all three samples fall off monotonically at large separations.

Figure \ref{fig:CMD_sep} displays the color-magnitude diagrams (CMD) for the disk-like (blue), intermediate (green) and halo-like (magenta) wide binaries. The unresolved close binary sequences \citep[ER18;][]{Widmark18} are visible but already sparse above the main sequence in the three subsamples. Following  ER18, we use the region at $\rm 1 \lesssim (G_{BP} - G_{RP}) \lesssim 2$ of the CMD (Figure~\ref{fig:CMD_sep}) to estimate the unresolved binary fraction of the disk and halo samples. In this region, the line ${\rm M_{G}=2.8\left(G_{BP}-G_{RP}\right)+2.4}$ divides the binary and single-star main sequences. We find that about 4.0\% (4.3\%), 10.9\% (7.4\%) and 6.6\% (4.8\%) of the primary (secondary) disk-like, intermediate and halo-like binaries probably have a bright unresolved companion. 

Figure \ref{fig:lb_distr} presents the final disk-like (the 4361 blue dots), intermediate (the 10,090 green dots) and halo-like (the 4351 magenta dots) binaries distribution in the $b$ vs. $l$ plane. Figure \ref{fig:hist_distrs} illustrates the distributions of distance ($\log(d)$, first column), magnitude difference ($\Delta \rm G=|G_1-G_2|$, second column), magnitude ($G$, third column), and angular separation ($\log\theta$, fourth column), respectively, for the three subsamples.  For demonstrating the contamination rate, the distributions of projected separations ($\log(s)$) for the three subsets are specifically displayed in Figure \ref{fig:hist_distr_sAU}. In this figure, the blue, green and magenta dotted curves show the distributions of the chance alignments selected from the shifted catalog with the disk-like, intermediate and halo-like criteria, respectively.

\begin{figure}[!t]
\centering
\includegraphics[width=0.4\textwidth, trim=0.0cm 0.0cm 0.0cm 0.0cm, clip]{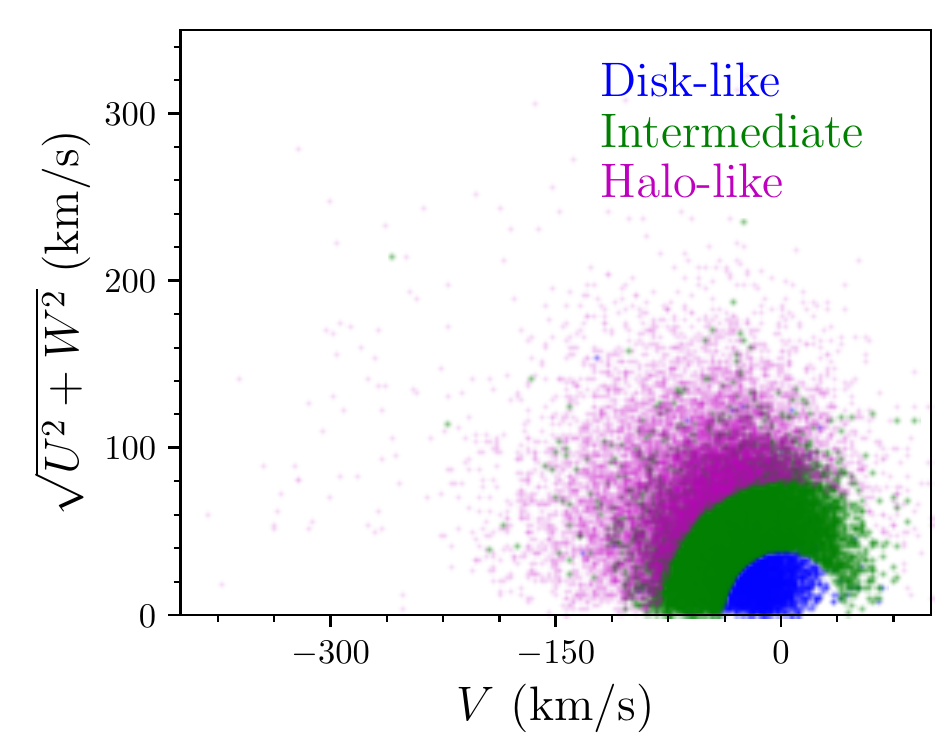}
\caption{Toomre diagram for disk-like (blue dots), intermediate (green dots) and halo-like (magenta dots) stars in the mock catalog separated using the cuts in $v_{\perp,\rm tot}$ that we use to select binaries in each kinematic sub-population. 
 }\label{fig:ToomreD}
\end{figure}

We use the mock Gaia DR2 catalog \citep{Rybizki2018} to illustrate the effect of $v_{\perp,\rm tot}$ for distinguishing the three subsamples. The mock catalog contains the full 6-D phase information for each star, so the disk-like, intermediate and halo-like stars are easily separated. We randomly select 150,000 stars in the whole sky from this mock catalog and use the criteria of $v_{\perp,\rm tot}$ to separate the mock stars into disk-like, intermediate and halo-like populations. Figure \ref{fig:ToomreD} shows the Toomre diagram for disk-like stars (blue points), intermediate (green points), and halo-like stars (magenta points). As one can see, the disk-like, intermediate and halo-like stars are clearly divided. This indicates that the criteria of $v_{\perp,\rm tot}$ work well for the selection of the three kinematic subsamples.

\section{Observed Separation Distributions}\label{sec:sep_distri}
We display distributions of the angular ($\theta$) and the projected physical separations ($s=(\theta \times (1/\varpi))$\,AU) for the disk-like, intermediate, and halo-like binaries in Figure \ref{fig:hist_distrs} (the last sub-panels) and Figure \ref{fig:hist_distr_sAU}. At $\theta<1.5$\,arcsec, there are no binaries in our sample, which is due to the Gaia angular resolution limit. At $1.5<\theta/\rm arcsec<6.0$ or $25<s/\rm AU<1000$ ($250<s/\rm AU<15000$ for the halo-like binaries), the samples are significantly incomplete. ER18 investigated this incompleteness effect and concluded that the incompleteness is due to blending of the two stars, which is more severe for binaries with magnitude difference between the two components. According to Figure A1 of ER18, the separation distribution is 84\% complete for $\Delta G < 5$ at $\theta > 5$\,arcsec and is complete for all $\Delta G$ at $\theta > 10$\,arcsec, {\it relative to the completeness at arbitrarily large angular separations}.

Consistent with ER18, we find that the main sequence binaries do not obey Opik's law (i.e., a flat distribution of log-separation) at any $s \gtrsim 500$\,AU. It is worthwhile to mention that there is an obvious drop at $s\sim10^{5.0}$\,AU in the raw separation distributions of the three subsamples, as shown in Figure \ref{fig:hist_distr_sAU}.

We note that, when insufficiently stringent astrometric quality cuts are applied, the separation distribution of all binary candidates in Figure~\ref{fig:hist_distr_sAU} appears bimodal (black histogram). However, the peak at large separations is due entirely to chance alignments, not real binaries. For this reason, it shifts toward wider separations and eventually disappears as more stringent quality cuts are applied. Some previous works \citep{Dhital_2010, Oelkers_2017} have found the separation distribution of wide binaries to be bimodal and have proposed that the wider-separation population contains binaries formed by different formation mechanism. Our analysis suggests that this apparent bimodality is simply due to chance alignments, as also suggested by \citealt{Andrews_2017}.

\section{Inferring Intrinsic Separation Distribution}
As discussed in Section \ref{sec:sep_distri}, the separation distribution at $\theta<10$\,arcsec is heavily incomplete. Before modeling the intrinsic-separation distribution, one must account for the selection effect to compensate for the incompleteness at small separation. In this section, we directly adopt an empirical fitting function from ER18 as the selection function, and we use a smoothly broken power law with four free parameters to model the intrinsic-separation distribution.

\subsection{Selection Function}
The selection function depends on the angular separation $\theta$ and the magnitude contrast $\Delta G$ between the two member stars. For this study, we empirically characterize the probability of detecting a companion at an angular separation $\theta$ with a fitting selection function $f_{\Delta G}(\theta)$ found by ER18,
 \begin{equation}
\label{eq:fitting_func}
f_{\Delta G}\left(\theta\right)=\frac{1}{1+\left(\theta/\theta_{0}\right)^{-\beta}}.
\end{equation}
Here $\Delta G$ is the magnitude contrast in $G$-band between the two member stars, $\theta_0$ characterizes the angular separation below which the sensitivity drops to 0, and $\beta$ determines how rapidly the sensitivity falls off at $\theta \ll \theta_0$. Following ER18, we adopt $\beta = 10$ for all $\Delta G$, $\theta_0 = 2.25$\,arcsec at $\Delta G < 1.5$\,mag, and $\theta_0 = 0.9(\Delta G + 1)$ at $\Delta G > 1.5$\,mag. We stress that $f_{\Delta G}(\theta)$ does not represent the absolute selection function, but rather the fraction of binaries detected at angular separation $\theta$ relative the the number that would be detected at arbitrarily large separation. See \citet{EBR2019b}, their Appendix D, for details.

\subsection{Function form of the separation distribution}
We model the separation distribution as a smoothly broken power law: 
\begin{equation}   
 \phi\left(s\right) =\phi_{0}\Bigl (\frac{s}{s_{\rm b}}\Bigr )^{\gamma_1} \Biggl [ \frac{1}{2}\Bigl [1+\Bigl (\frac{s}{s_{\rm b}}\Bigr )^{1/\Lambda}\Bigr ] \Biggr ]^{(\gamma_2 - \gamma_1)\Lambda}
 \label{eq:b_power}           
 \end{equation}
where $\phi_0$ is a normalization parameter. The break separation $s_{\rm b}$ marks the transition between the two single power laws with index $\gamma_1$ and $\gamma_2$, respectively. The $\Lambda$ parameter quantifies how abruptly or smoothly the two power laws are joined. The model has four fit parameters $\vec{m} = (\gamma_1, \gamma_2, \log(s_{\rm b}/{\rm AU}), \Lambda)$. We suppose that $\phi(s)$ is independent of both the distance and the absolute magnitude of the two member stars, although the separation distribution {\it is} expected to vary somewhat with the mass (and thus, absolute magnitude) of the two stars \citep[e.g.,][]{Duchene_2013, MD2017, MK2018}.  This supposition is validated with two special datasets selected from the pure samples: one with $\Delta G>2.5$\,mag, which contains 1633 disk-like, 3481 intermediate, and 986 halo-like binaries; the other one with $M_{G1}<7$\, mag, which includes 2287 disk-like, 6684 intermediate, and 3542 halo-like binaries. The analysis on these datasets indicates that the large $\Delta G$ or stellar mass just slightly affects the shape of the observed separation distribution at small separations, and this selection effect can be well overcome with the selection function when the intrinsic separation distribution is derived in the following section.

\subsection{Likelihood for Fitting the Separation Distribution}
Given a sample of binaries with projected separations $s_i$, the likelihood function is
 \begin{equation}
L=p\left(\left\{ s_{i}\right\} |\vec{m}\right)=\prod_{i}p\left(s_{i}|\vec{m}\right),
\end{equation}
where $\vec{m}$ is a set of free model parameters to be fitted and $p(s_{i}|\vec{m})$ is the probability of detecting the $i$-th binary given model parameters $\vec{m}$. For the $i$-th binary, $p(s_{i}|\vec{m})$ can be specified
 \begin{equation}
p\left(s_{i}|\vec{m}\right)=\frac{\phi\left(s_{i}|\vec{m}\right)}{\int_{s_{{\rm min}}}^{s_{{\rm max}}}\phi\left(s|\vec{m}\right)f_{\Delta G}\left(s|d_{i}\right)\,{\rm d}s}
\label{eq:p_i}
\end{equation}
where $\phi\left(s_{i}|\vec{m}\right)$ is the probability that a binary with distance $d_i$, magnitude difference $\Delta G$, and physical separation $s_i$ is found in the catalog, given an intrinsic-separation distribution (specified by Equation~(\ref{eq:b_power})) with parameters $\vec{m}$. The selection function, $f_{\Delta G}\left(s/d_{i}\right)=f_{\Delta G}\left(\theta\right)$, is given by Equation~(\ref{eq:fitting_func}). Here $s_{\rm min}$ and $s_{\rm max}$ are the minimum and maximum separations of the observed distribution in the disk and halo binary samples, and $\phi(s|\vec{m})$ is normalized such that $\int_{s_{{\rm min}}}^{s_{{\rm max}}}\phi\left(s|\vec{m}\right)\,{\rm d}s=1$. Equation~(\ref{eq:p_i}) does not account for the observational uncertainties in $s_i$ or $d_i$. These are small because all the binaries in our clean samples have parallax errors smaller than 5\%. 

The denominator in Equation~(\ref{eq:p_i}) reflects the fraction of predicted binaries that could have been detected at a distance $d_i$ and magnitude difference $\Delta G$; it accounts for the fact that at large distances and large $\Delta G$, only binaries with large $s$ can be detected.  We set $s_{\rm max} = 10^{5.25}$\,AU (i.e., $\sim 0.8$\,pc),  omitting the widest separation bin due to a high contamination rate beyond $10^{5.25}$\,AU. The choice of $s_{\rm min}$ has no effect on our results because the integrand in Equation~(\ref{eq:p_i}) goes to 0 at small separations. We set $s_{\rm min} = 10^{-2}$\,AU. 

If either member star of a binary is too faint, the binary probably can not be detected. Therefore, whether a binary can be observed or not also depends on the apparent magnitude of both stars. However, this has no effect on the inferred separation distribution as long as the undetected binaries have the same intrinsic separation distribution.

\section{Results}\label{sect:result}
\begin{figure*}[!t]
\centering
\includegraphics[width=0.75\textwidth, trim=0.0cm 0.0cm 0.0cm 0.0cm, clip]{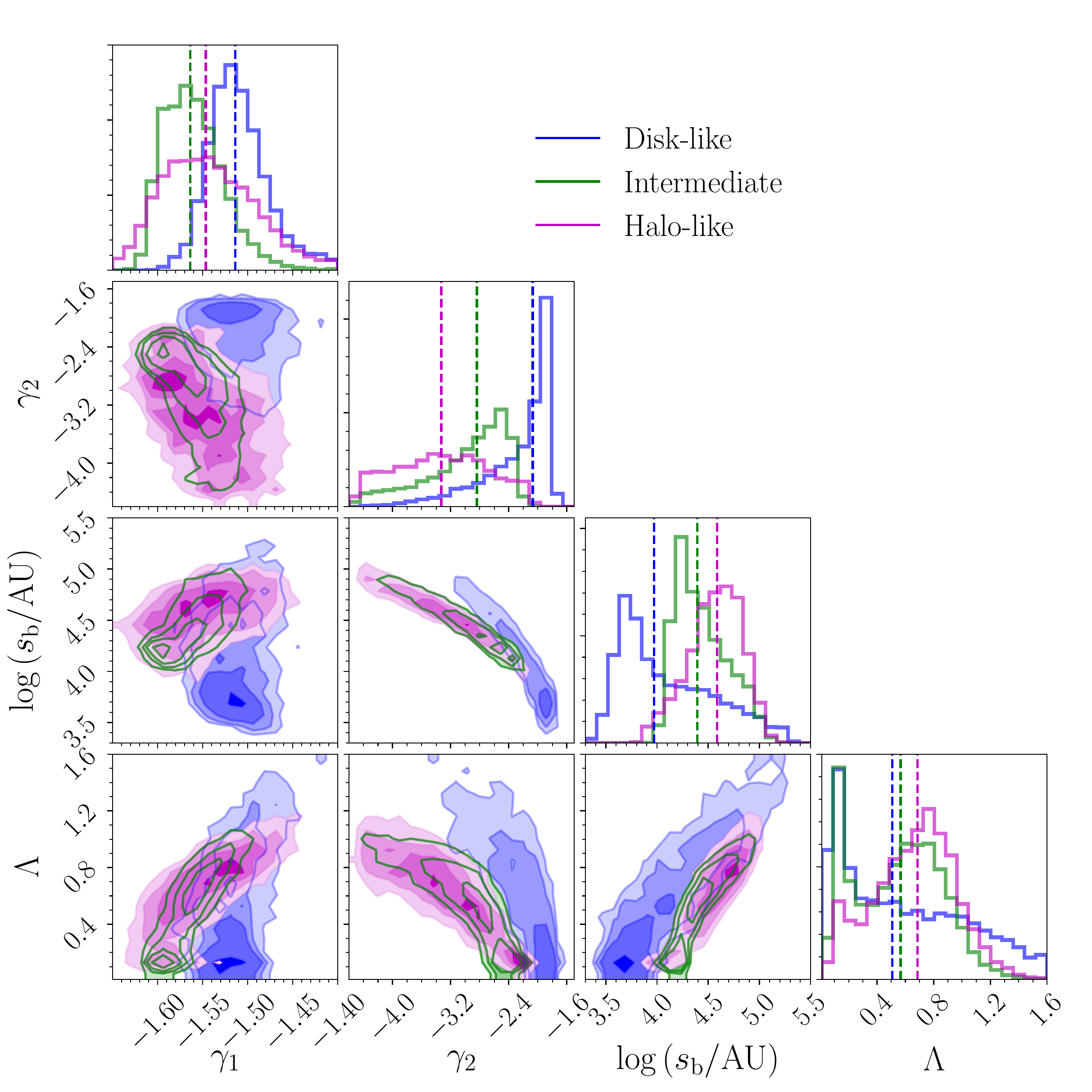}
\includegraphics[width=0.75\textwidth, trim=0.0cm 0.0cm 0.0cm 0.0cm, clip]{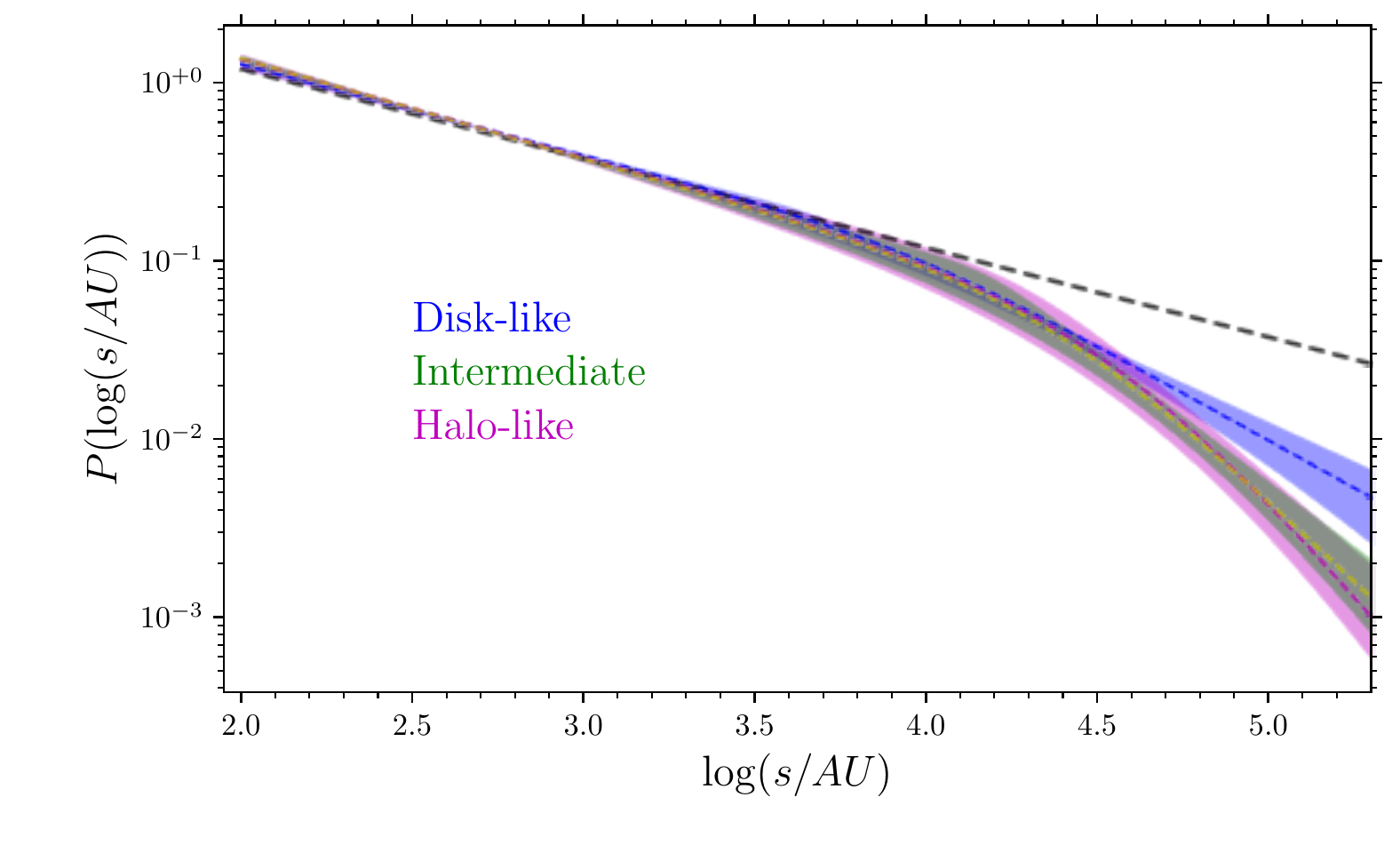}
\caption{Top: contours and marginalized probability distributions of the four parameters of the smoothly broken power law for the intrinsic-separation distribution of the disk-like (blue), intermediate (green) and halo-like (magenta) binaries, visualized with corner.py \citep{corner}. The dashed lines mark the best-fit constraints for each sample. Bottom: intrinsic distributions of projected separation for each sample. The uncertainties (within $1\sigma$) are displayed with shaded regions. The separation distributions of the three subsamples are indistinguishable at small separations, $s<10^{4.4}$\,AU ($\sim0.12$\,pc). The separation distributions for all three subsamples deviate from a single power law (illustrated with the black dashed line) at $s>10^{4.0}$\,AU ($\sim0.05$\,pc), with a somewhat steeper fall-off at wide separations for halo binaries than for disk binaries. 
} \label{fig:fitting}
\end{figure*}

We use \texttt{emcee} \citep{FormanMackey_2013} to sample the posterior distribution of the broken-power-law model of binary separations (Eq.\ref{eq:b_power}), for the three subsamples of disk-like, intermediate and halo-like binaries. We use flat priors for $\gamma_1$, $\gamma_2$, and $\log(s_{\rm b}/{\rm AU})$ and use an exponential prior on $\Lambda$, taking a prior of $\exp(\textendash3\Lambda$) for $\Lambda$. The top panel of Figure \ref{fig:fitting} displays the contours and marginalized probabilities of the model parameters posteriors $\vec{m}$, while the bottom panel illustrates the inferred intrinsic separation distributions for the three subsamples. The median values of $\vec{m} =(\gamma_1, \gamma_2, \log(s_{\rm b}/{\rm AU}),\Lambda)$ are

\begin{itemize}
    \item $(-1.51^{+0.03}_{-0.03}, -2.07^{+0.69}_{-0.19}, 3.97^{+0.67}_{-0.32}, 0.51^{+0.59}_{-0.39})$ for the disk-like binaries,
    \item $(-1.56^{+0.03}_{-0.04}, -2.84^{+0.77}_{-0.38}, 4.39^{+0.34}_{-0.21}, 0.57^{+0.29}_{-0.41})$ for the intermediate binaries
    \item $(-1.55^{+0.05}_{-0.06}, -3.33^{+0.72}_{-0.68}, 4.59^{+0.24}_{-0.29}, 0.67^{+0.26}_{-0.34})$ for the halo-like binaries. 
\end{itemize}

These results, which are illustrated in Figures \ref{fig:fitting}, can be summarized as follows:
  \begin{enumerate}
    \item At $\rm s<s_b$, the three subsamples have the same separation distributions, i.e., a power law with an index of $\gamma_1\sim -1.54$ shown as the black dashed lines in the bottom panels of Figure \ref{fig:fitting}. 
      \item At $\rm s>s_b$, the separation distributions become significantly steeper for all three subsamples, excluding single power-law models over the entire separation range (see the bottom panel of Figure \ref{fig:fitting}). This profile steepening ($\Delta \gamma\equiv \gamma_2 - \gamma_2$) increases from 0.56$\pm$0.50 for the disk-like binaries, through $1.28\pm$0.60 for the intermediate binaries, to 1.78$\pm$0.70 for the halo-like binaries.  There is thus a strong hint of a steeper fall-off at ultra-wide separations ($s>10^{4.5}$\,AU) for the halo-like sample than the disk sample, although the marginalized constraints on $\gamma_2$ are quite broad due to degeneracies with other parameters.
        \item The $\Delta \gamma$ for the disk-like sample is significant, but small enough to explain why ER18 found comparably good fits for MS/MS binaries with a single power law with a slope -1.6 over $500<\rm s/AU<50,000$. The $\Delta \gamma = 1.78\pm0.70$ for the halo-like sample, however, indicates a quite substantive change in the separation distribution slope at $s\sim10^{4.3}$\,AU (i.e., $\sim$0.1\,pc).
  
    \item The smoothing parameter $\Lambda$ differs among the subsamples: the halo-like subsample has the strongest break (highest $\Delta\gamma$), yet the smoothest power-law transition, i.e. the highest parameter $\Lambda$. 
        
    \item  The posterior distributions of the free parameters (except $\gamma_1$) gradually present a double-peak feature from the disk-like to the halo subsamples. It indicates that the intrinsic separation distributions favor two breaks for all the three subsamples, in particular for the old population. One break takes place at $s\sim10^{3.8}$\,AU, the corresponding smoothing parameter is $\Lambda\sim0.05$. The other break is at $s\sim10^{4.5}$\,AU and its $\Lambda\sim0.7$.
    
    \item The three free parameters, i.e., $\gamma_2$, $\log(s_{\rm b}/{\rm AU})$ and $\Lambda$, are degenerate and correlated with each other. So we will discuss the two breaks with two fixed $\Lambda$ , i.e., $\Lambda=0.05$ (Figure \ref{fig:fitting2}) and $\Lambda=0.7$ (Figure \ref{fig:fitting3}), respectively, in the Appendix. 
             
\end{enumerate}

\section{Interpreting Separation Distributions\label{sect:theory}}

At smaller separations, the observed separation distribution of binaries presumably reflects a combination of birth separations and subsequent evolution \citep[e.g.][and references therein]{EBR2019b, liu2019, MD2017}. For the very wide binaries (beyond a few $1000$ AU) of primary interest here, the formation process leading to the "initial" separation distribution at the time of birth cluster dispersal, and the subsequent evolutionary processes, are not firmly established. Wide binaries presumably did not form within a common 
disk, but resulted from turbulent fragmentation (at closer separations) and stochastic pairing during cluster dissolution (at wider separations).
After formation, wide binaries could retain their original orbital parameters, except (1) that they may be affected or become disrupted by gravitational encounters with molecular clouds or other massive objects \citep[e.g.][]{CG04}; (2) that either one or both binary members evolve off main sequence, inducing rapid mass loss (e.g. ER18); or (3) that dynamical evolution leads to widening within unstable triples \citep[e.g.]{Reipurth_2012}.

This present work was originally initiated with the goal of detecting or constraining signs of binary disruption by massive objects. In such an analysis one must presume to have confident prior knowledge about the initial separation distribution to $\sim 1$~pc, e.g. that it was a perfect single power-law at large $s$. We first pursue this approach of modeling, asking what population of massive scattering objects could lead to the breaks seen. This, however, turns out to have physically implausible implications, at variance with other constraints. In light of this we then ask what separation distribution we may have {\it expected} at birth, and find it to be consistent with the observations.

\subsection{An initial single power law, broken \\ by a population of scatters?}
We start out by noting that the outer slope of the disk-like binary is if anything flatter than that of the halo-like populations, showing a steepening by only $\lesssim 0.6$: this argues against, or at least provides not evidence for, molecular clouds being a dominant mechanism is setting the outer slope of wide disk-like binaries by means of tidal disruption that would induce a cut-off.

We  now use the approach of previous studies, assuming that the separation distribution for wide binaries follows a single (initial) power law. By implication, any break in such power law is then attributed 
to external perturbations, e.g., binary disruption as a consequence of encountering MACHOs. 
Disk-like binaries are more likely to encounter molecular clouds - acting as perturbers -  during much of their orbits, and any such encounters will be at lower velocities. It has therefore been inferred \citep[e.g.][]{CG04} that such samples should not be used to probe the properties of MACHOs, as their perturbative effect would be subdominant. However, halo binaries spend only a small fraction of their lifetime in the disk and cross it at high speed, so MACHOs (if they exist) could be the dominant perturbers, with the observed separation distribution providing constraints on their density and mass. 

Such external perturbations can be usefully divided into two regimes: a diffusive regime in which perturbations are described by a multitude of individually weak gravitational encounters, and a catastrophic regime in which perturbations are dominated by a single closest encounter. As described in Section \ref{sect:result}, we detect steepening at $\sim0.1$\,pc in the separation distribution, which is particularly distinct for the halo-like subsample. 

For those two disruption regimes \citep[Eq. 8.65a and b]{BT2008}, the timescales to disrupt a solar mass binary at a separation of $\sim 0.1$~pc can be expressed as:
\begin{equation}\label{eq:dv12}
t_{d, \rm diff} \approx 3~{\rm Gyr} \Bigl [ \frac{v}{200\rm km/s}\Bigr ]~\Bigl [\frac{30\rm M_{\odot}}{\rm M}\Bigr ]~\Bigl [ \frac{\rho_H}{\rho}\Bigr ]~\Bigl [\frac{0.1\rm pc}{s}\Bigr ]
\end{equation}
and 
\begin{equation}\label{eq:dv12a}
t_{d, \rm cat} \approx 3~{\rm Gyr} \Bigl [ \frac{\rho_H}{\rho}\Bigr ]~\Bigl [\frac{0.1\rm pc}{s}\Bigr ]^{3/2}.
\end{equation}
Here, $v$ is the relative velocity of a MACHO perturber passing by one of binary components, $M$ is the individual mass of the perturbers, and $\rho$ is their mean mass density near the Sun. The local halo mass density is denoted by $\rho_H$, and we adopted $\rho_H=0.01$\Msun pc$^{-3}$. In the catastrophic regime, the disruption timescale at a given separation is independent of the mass and velocity of the perturber, set merely by $\rho$. If the perturber mass
\begin{equation}\label{eq:pho}
M>10\,M_{\odot},
\end{equation}
$t_{d, \rm diff}$ would be shorter than the typical time, i.e., 10\,Gyr.
The condition to see a distinct break due to disruption in the diffusive regime is $\frac{30\rm M_{\odot}}{\rm M}\cdot \frac{\rho_H}{\rho}~<~1$. Therefore, $M_{\rm crit}\sim 30$\,\Msun, marks the transition between the two regimes. Previous authors expected to see a signature of binary separation function at $s\sim0.1$\,pc in the several studies \citep[CG04; Y04;][]{Quinn09}, but failed to detect the break signature due to the paucity of observational wide binaries at large separations.

However, the implied {\sl lower} limits on the MACHO density or MACHO mass implied by the above analysis (critically hinging on the {\it assumption} of an initial single power law for the binary separations) would be in direct conflict with the the stringent {\sl upper} limits from the survival of compact ultra-faint dwarf galaxies and the star cluster in Eridanus II. With these latter systems, \citet{Brandt2016} closed the window of allowed MACHO mass in $\sim20-100$\Msun, and thereby ruled out the entire window of MACHO mass in $>10^{-7}$\Msun\ by combining his results with the existing constraints \citep[Y04;][]{Tisserand07,Wyrzykowski08,Quinn09,Green_2016}. We also note that the steepening in the observed separation distributions is relatively gradual, and is not manifest as a sharp truncation or break. 
The arguably most obvious way to resolve this tension is to acknowledge that the assumption of an initial single power law was too restrictive, and not realized in nature.

\subsection{A simple model for the primordial\\ wide-binary separation distribution }

We now lay out a very simple model for the separation distribution of wide binaries (wider than a few 1000~AU) that might be expected as a consequence of their formation. Given their dynamical fragility, we presume that most could have formed only as the cluster (or association) was dispersing. 
A pair of stars in a dispersing star cluster at separation $s$ will be and presumably remain
bound provided that $v_\tot < v_\max(s,m) = \sqrt{2Gm_\tot/s}$, where
$m_\tot$ is the total mass of the pair and
$v_\tot$ is their total relative velocity. For a Gaussian velocity
distribution $\sigma$ (the velocity dispersion of the birth cluster),
and for ``wide binaries'' ($v_\tot\ll \sigma$),
this will occur with probability,
\begin{equation}\label{eq:probwide}
p(s~|~m_\tot,\sigma ) = {4\pi\over 3}\biggl({v_\tot\over \sqrt{8\pi}\sigma}\biggr)^3
  \simeq  0.1 G^{3/2}\sigma^{-3}\times m_\tot^{3/2}s^{-3/2},
\end{equation} 
for separations smaller than the cluster's tidal radius, $R_\tidal$. 
Here we have taken into account the fact that while phase space
available to yield $s$ scales $\propto s^2 ds$, the $s^2$ in this expression is
canceled for a typical cluster profile $\rho\propto r^{-2}$,
within $R_\tidal$, with no pairs larger than this radius.

Thus for an ensemble of dissolving clusters with characteristic
tidal radii $R_\tidal \sim {\cal O}(R_\typ)$, the separation distribution
of subsequently bound wide pairs for $s\ll R_\typ$ will always be the same (i.e., $s^\gamma$, 
$\gamma=-3/2$); the sum of such distribution will
have the same power law.  However, as these clusters will not
all have the same $R_\tidal$, the individual-cluster
distributions will have breaks at $s\sim R_\typ$.  Hence, for $s\ll R_\typ$, the joint distribution will
be characterized by a true power law, while for $s\ga R_\typ$
it will fall off more rapidly in a manner that presumably could be approximated
as a power law, depending on the statistics of $R_\typ$. 

This simple model has two attractive features in this context. First, it provides an explanation of why wide binaries in the range $\sim 1000$AU to $10,000$~AU have a separation distribution of $\gamma\eqsim -1.5$ across all populations. And it implies that the birth cluster size has an imprint on the (maximal) binary separation distribution even long after the cluster has been dispersed. If stars born in earlier epochs of the Milky Way's evolution ({\it viz.} the halo-like population) formed in clusters that were more compact (smaller $R_\tidal$), or with a narrower distribution of $R_\tidal$ among the clusters, then their separation distribution should experience a more distinct departure from the $\gamma\simeq -1.5$ power law. On the other hand, halo binaries are subject to fewer gravitational perturbations due to their higher velocities. These two effects work in opposite directions and it is not immediately clear which will dominate. The fact that the observed separation distributions across the three samples are quite similar even at wide separations suggests that they partially cancel. If the clusters that gave birth at a later epoch to the disk-like binaries had a wider distribution of $R_\tidal$, e.g. because a larger fraction of them were born in barely-bound associations, then we would expect a less distinct break in the separation power-law.
In this model, the present-day separation distribution is still mostly a reflection of the initial distribution.
Of course, some tidal disruption may have played a role as $s$ approaches one parsec.
Note also, that this model can only infer - not predict - the break radius, $s_b$, which is larger for the halo-like binaries than for the disk-like ones.

\section{Conclusion}\label{sect:conclution}
We have compiled an extensive yet pure catalog of candidate wide binaries ($a<1$~pc) selected from Gaia DR2 in the solar neighborhood with distances $d<4$~kpc, following a procedure similar to that of ER18. This initial candidate catalog consists of 807,611 possible binaries. Its contamination rates are lower than 10\% at $a<20,000$~AU; however, the contamination rates quickly increase beyond 20,000~AU, until up to 100\% at the largest separation bin, i.e., $a\sim 1.0$\,pc. To address this, we subsequently applied additional selection criteria, tailored towards three kinematically-selected, presumably pure subsamples: disk-like, intermediate, and halo-like binaries.  The raw catalog is described in Table \ref{tab:catalog}, which will be released on-line {\bf and available via the PaperData Service of China-VO}.

This raw catalog of candidate wide binaries can provide a starting point for several scientific applications beyond the scope of this paper. For instance, one could select a subsample of wide binaries containing a WD member from the raw catalog to determine stellar ages. The age of a WD can be easily constrained from its cooling age and mass given an initial-final mass relation and an initial mass-age relation, while the age of its companion MS star could not be measured precisely. The two members in a binary usually can be regards to be co-eval, so the MS star's age can be simply obtained from its companion WD star \citep{Fouesneau2019}. 

To reach low contamination to separations of $\sim 1$~pc, we defined three subsamples of MS-MS binaries from the raw catalog via the systems' projected velocities with respect to the Sun, $v_{\perp,tot}$: 4361 disk-like binaries with $v_{\perp,tot}\le 40$~km/s, 10,090 kinematically intermediate binaries with $40\le $~km/s$\le v_{\perp,tot}\le 85$~km/s, and 4351 binaries with halo-like kinematics ($v_{\perp,tot}\ge 85$~km/s). We presume that these velocity cuts represent a rough ordering in the binaries' age. Using three catalogs, i.e., the mock Gaia DR2 catalog \citep{Rybizki2018}, a shifted Gaia DR2 catalog, and around 7.2 million Gaia DR2 stars with RVs,  we quantify the contamination rates of the three subsamples after we have applied additional cuts in the different separation bins, and find that the contamination rates in the three subsamples are negligible in the range of separation $s< 1.0$\,pc. 

We then proceeded to model the separation distributions of these subsamples as a smoothly broken power law, with four free parameters $\vec{m} = (\gamma_1, \gamma_2, \log(s_{\rm b}/{\rm AU}), \Lambda)$. Fitting these model for the three subsamples for $s<10^{5.3}$~AU and accounting for the selection function, several important findings have been discovered: (1) we confirm that the slope for separations from $10^{2.5}$~AU to $10^{4.0}$~AU is $p(s)\propto s^\gamma = s^{-1.54}$, as found in previous studies; this slope is essentially the same for all three subsamples. (2) we show for the first time at high significance how the slope steepens beyond $10^4~\rm AU\le s\le 10^5$~\rm AU; and we find first tentative evidence that this slope-steepening differs among these sub-populations: by only $\Delta\gamma\sim 0.5$ for stars of disk-like kinematics, but by $\Delta\gamma=1-3$ for stars with halo-like kinematics. The actual values of $\Delta\gamma$ would be even larger than these estimates, if we had unrecognized contamination at the largest separations. (3) we also find some, albeit tentative, evidence that $s_b$ increases from $5,000$~AU for the disk-like subsample to $20,000$~AU for the halo-like subsample. 

We have offered interpretations for these observational findings in the two limiting cases: we start by presuming
that the birth separation was a single power-law to $\sim 1$~pc, altered only by binary disruption effected by some 
possible MACHO populations. The break in $\sim0.1$\,pc in the separation distribution of the halo-like subsample, 
would imply $M>10$~\Msun\, at a very high mean MACHO density. However, this window of MACHO candidates 
has been closed by other studies, e.g., \citet{Brandt2016}, and we must conclude that -- in light of a complex, but not truncated, separation distributions -- MACHO limits or detections would be hard to derive, as no clearly defined null-hypothesis for the undisturbed separation distribution exists.

We therefore build a conceptually different simple model asking what we should expect for the initial (birth) separation distribution. We find that a simple model, where binaries form by stars remaining stochastically bound as the cluster disperses, works remarkably well in two respects: in generically predicts a distribution power law of $s^{-\frac{3}{2}}$, as observed in all populations. And it implies that the separation distribution will steepen beyond separations in excess of the initial cluster size; therefore, (expected) structural differences between the birth clusters or associations of the different populations may now be reflected in the separation distribution of ultra-wide binaries.  

\acknowledgements
The authors thank Bing Yue, Qiang Yuan, Chao Liu and Ling Zhu for the helpful discussions. HJT acknowledges the National Natural Science Foundation of China (NSFC) under grants 11873034, U1731124. KE was supported in part by an NSF graduate research fellowship and by SFB 881. This work has made use of data from the European Space Agency (ESA) mission
{\it Gaia} (\url{https://www.cosmos.esa.int/gaia}), processed by the {\it Gaia}
Data Processing and Analysis Consortium (DPAC, \url{https://www.cosmos.esa.int/web/gaia/dpac/consortium}). Funding for the DPAC
has been provided by national institutions, in particular the institutions
participating in the {\it Gaia} Multilateral Agreement.

\begin{table*}
\centering
\caption{Catalog description}
\label{tab:catalog}
\begin{tabular}{p{4.0cm} | p{1.cm}| p{11.5cm} } 
\hline
\hline
Column & units & Description \\
\hline
\texttt{source\_id}                          &         & {\it Gaia} source id (int64); star 1  \\
\texttt{source\_id2}                         &         & {\it Gaia} source id (int64); star 2  \\
\texttt{astrometric\_chi2\_al}               &         & astrometric goodness-of-fit ($\chi^2$) in the along-scan direction; star 1   \\
\texttt{astrometric\_chi2\_al\_2}            &         & astrometric goodness-of-fit ($\chi^2$) in the along-scan direction; star 2  \\
\texttt{astrometric\_n\_good\_obs\_al}       &         & number of good CCD transits; star 1 \\
\texttt{astrometric\_n\_good\_obs\_al2}      &         & number of good CCD transits; star 2  \\
\texttt{dec}                                 & deg     & declination; star 1  \\
\texttt{dec2}                                & deg     & declination; star 2  \\
\texttt{ra}                                  & deg     & right ascension; star 1  \\
\texttt{ra2}                                 & deg     & right ascension; star 2  \\
\texttt{pairdistance}                        & deg     & angular separation between star 1 and star 2  \\
\texttt{parallax}                            & mas     & parallax; star 1  \\
\texttt{parallax2}                           & mas     & parallax; star 2  \\
\texttt{parallax\_over\_error}               &         & parallax divided by its error; star 1  \\
\texttt{parallax\_over\_error2}              &         & parallax divided by its error; star 2 \\
\texttt{phot\_bp\_mean\_flux\_over\_error}   &         & integrated BP mean flux divided by its error; star 1  \\
\texttt{phot\_bp\_mean\_flux\_over\_error2}  &         & integrated BP mean flux divided by its error; star 2  \\
\texttt{phot\_bp\_mean\_mag}                 & mag     & integrated BP mean magnitude; star 1  \\
\texttt{phot\_bp\_mean\_mag2}                & mag     & integrated BP mean magnitude; star 2  \\
\texttt{phot\_bp\_rp\_excess\_factor}        &         & ratio of total integrated BP and RP flux to G-band flux; star 1 \\
\texttt{phot\_bp\_rp\_excess\_factor2}       &         & ratio of total integrated BP and RP flux to G-band flux; star 2  \\
\texttt{phot\_g\_mean\_flux\_over\_error}    &         & integrated G-band mean flux divide by its error; star 1  \\
\texttt{phot\_g\_mean\_flux\_over\_error2}   &         & integrated G-band mean flux divide by its error; star 2  \\
\texttt{phot\_g\_mean\_mag}                  & mag     & G-band mean magnitude (Vega scale); star 1  \\
\texttt{phot\_g\_mean\_mag2}                 & mag     & G-band mean magnitude (Vega scale); star 2  \\
\texttt{phot\_rp\_mean\_flux\_over\_error}   &         & integrated RP mean flux divided by its error; star 1  \\
\texttt{phot\_rp\_mean\_flux\_over\_error2}  &         & integrated RP mean flux divided by its error; star 2  \\
\texttt{phot\_rp\_mean\_mag}                 & mag     & integrated RP mean magnitude; star 1  \\
\texttt{phot\_rp\_mean\_mag2}                & mag     & integrated RP mean magnitude; star 2  \\
\texttt{pmdec}                               & mas\,yr$^{-1}$ & proper motion in the declination direction; star 1  \\
\texttt{pmdec2}                              & mas\,yr$^{-1}$ & proper motion in the declination direction; star 2  \\
\texttt{pmdec\_error}                        & mas\,yr$^{-1}$ & standard error of proper motion in the declination direction; star 1  \\
\texttt{pmdec\_error2}                       & mas\,yr$^{-1}$ & standard error of proper motion in the declination direction; star 2  \\
\texttt{pmra}                                & mas\,yr$^{-1}$ & proper motion in right ascension direction; i.e., $\mu_{\alpha}^{*} = \mu_{\alpha}\cos{\delta}$; star 1  \\
\texttt{pmra2}                               & mas\,yr$^{-1}$ & proper motion in right ascension direction; i.e., $\mu_{\alpha}^{*} = \mu_{\alpha}\cos{\delta}$; star 2  \\
\texttt{pmra\_error}                         & mas\,yr$^{-1}$ & standard error of proper motion in right ascension direction; star 1  \\
\texttt{pmra\_error2}                        & mas\,yr$^{-1}$ & standard error of proper motion in right ascension direction; star 2  \\
\texttt{radial\_velocity}                    & km\,s$^{-1}$ & spectroscopic barycentric radial velocity; star 1  \\
\texttt{radial\_velocity2}                   & km\,s$^{-1}$ & spectroscopic barycentric radial velocity; star 1  \\
\texttt{radial\_velocity\_error}             & km\,s$^{-1}$ & standard error of spectroscopic barycentric radial velocity; star 1  \\
\texttt{radial\_velocity\_error2}            & km\,s$^{-1}$ & standard error of spectroscopic barycentric radial velocity; star 2  \\
\texttt{rv\_nb\_transits}                    &         & number of transits used to compute radial velocity; star 1  \\
\texttt{rv\_nb\_transits2}                   &         & number of transits used to compute radial velocity; star 2  \\
\texttt{s\_AU}                               & AU      & projected physical separation between two stars \\
\texttt{binary\_type}                               &       & type of wide binary. Different types are assigned different integer identifiers: MS-WD, WD-WD, and MS-MS are assigned with 1, 2, and 3, respectively. \\
\texttt{num}                               &       & number of pairs in neighborhood for each binary\\
\texttt{U}                    & km\,s$^{-1}$ & radial velocity in the Cartesian coordinates (positive to the Galactic Center); star 1  \\
\texttt{V}                   & km\,s$^{-1}$ & azimuthal velocity in the Cartesian coordinates; star 1  \\
\texttt{W}             & km\,s$^{-1}$ & vertical velocity in the Cartesian coordinates; star 1  \\
\texttt{U2}                    & km\,s$^{-1}$ & radial velocity in the Cartesian coordinates; star 2  \\
\texttt{V2}                   & km\,s$^{-1}$ & azimuthal velocity in the Cartesian coordinates; star 2  \\
\texttt{W2}             & km\,s$^{-1}$ & vertical velocity in the Cartesian coordinates; star 2  \\
\texttt{U\_error}                    & km\,s$^{-1}$ & error of radial velocity in the Cartesian coordinates; star 1  \\
\texttt{V\_error}                   & km\,s$^{-1}$ & error of azimuthal velocity in the Cartesian coordinates; star 1  \\
\texttt{W\_error}             & km\,s$^{-1}$ & error of vertical velocity in the Cartesian coordinates; star 1  \\
\texttt{U\_error2}                    & km\,s$^{-1}$ & error of radial velocity in the Cartesian coordinates; star 2  \\
\texttt{V\_error2}                   & km\,s$^{-1}$ & error of azimuthal velocity in the Cartesian coordinates; star 2  \\
\texttt{W\_error2}             & km\,s$^{-1}$ & error of vertical velocity in the Cartesian coordinates; star 2  \\
\hline
\hline
\end{tabular}
\begin{flushleft}
{\bf Note}: Each row in the catalog corresponds to a single binary; ``star 1'' and ``star 2'' designations in each binary are arbitrary. Full descriptions of {\it Gaia} variables can be found at \Gaiaurl.
\end{flushleft}
\end{table*}

\begin{figure}[!t]
\centering
\includegraphics[width=0.7\textwidth, trim=0.0cm 0.0cm 0.0cm 0.0cm, clip]{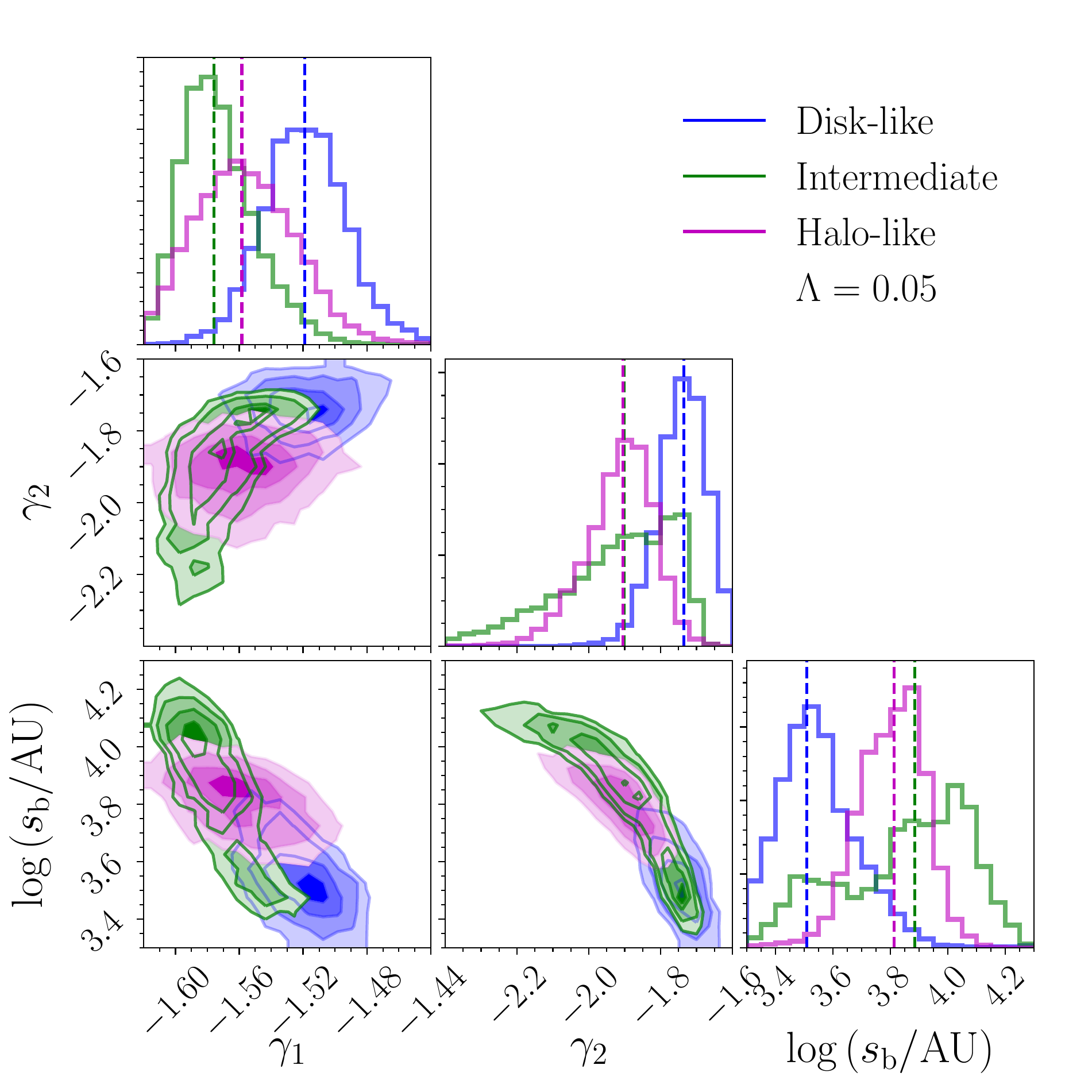}
\includegraphics[width=0.7\textwidth, trim=0.0cm 0.0cm 0.0cm 0.0cm, clip]{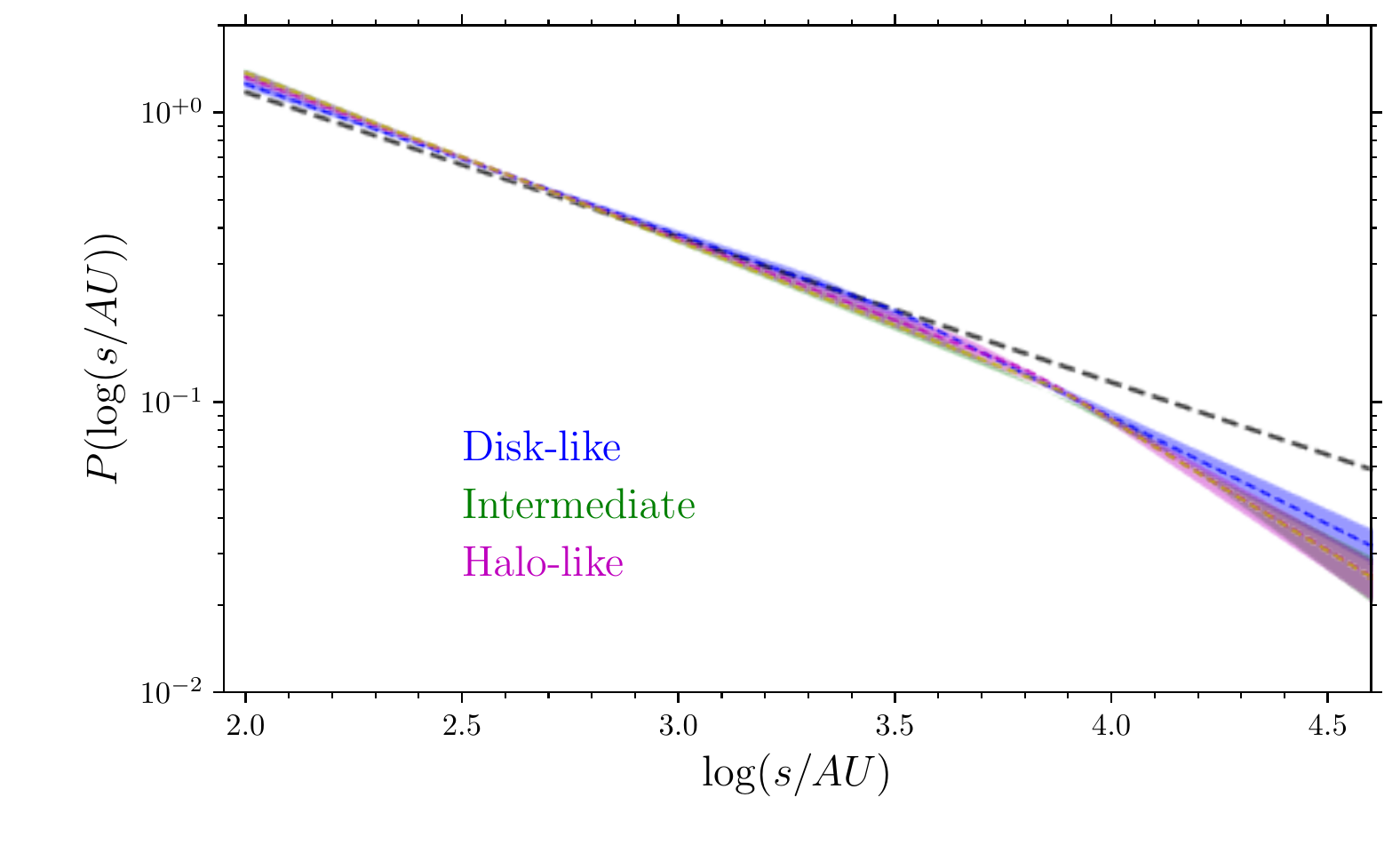}
\caption{Similar to Figure \ref{fig:fitting}, but with a fixed smoothing parameter, i.e., $\Lambda=0.05$, and setting $s_{\rm max}=10^{4.6}$~AU. This figure highlights the small-separation break in the inferred intrinsic separation distribution, which is hard to identify from the observed separation distributions in Figure \ref{fig:hist_distr_sAU}. At $s<10^{4.5}$\,AU, the distributions for the three samples are formally consistent, although there is a hint of a steeper fall-off at large separations for the halo-like sample than the disk sample.
 }\label{fig:fitting2}
\end{figure}

\appendix \label{app:A}
In order to quantify and clearly illustrate the smaller-separation break in $p(s)$, we sample the posterior distribution of the broken-power-law model with a fixed value of $\Lambda=0.05$, setting $s_{\rm max}=10^{4.6}$~AU for each binary subsample. The resulting constraints on the intrinsic separation distributions are displayed in Figure \ref{fig:fitting2}, similar to Figure \ref{fig:fitting}. The best fit values of $\vec{m} =(\gamma_1, \gamma_2, \log(s_{\rm b}/{\rm AU}))$ are
\begin{itemize}
    \item $(-1.52^{+0.03}_{-0.03},-1.74^{+0.07}_{-0.06}, 3.51^{+0.16}_{-0.19})$ for the disk-like binaries,
    \item $(-1.58^{+0.02}_{-0.03}, -1.90^{+0.21}_{-0.14}, 3.88^{+0.18}_{-0.33})$ for the intermediate binaries
    \item $(-1.56^{+0.03}_{-0.03}, -1.90^{+0.11}_{-0.08}, 3.81^{+0.10}_{-0.13})$ for the halo-like binaries. 
\end{itemize}

\begin{figure}[!t]
\centering
\includegraphics[width=0.7\textwidth, trim=0.0cm 0.0cm 0.0cm 0.0cm, clip]{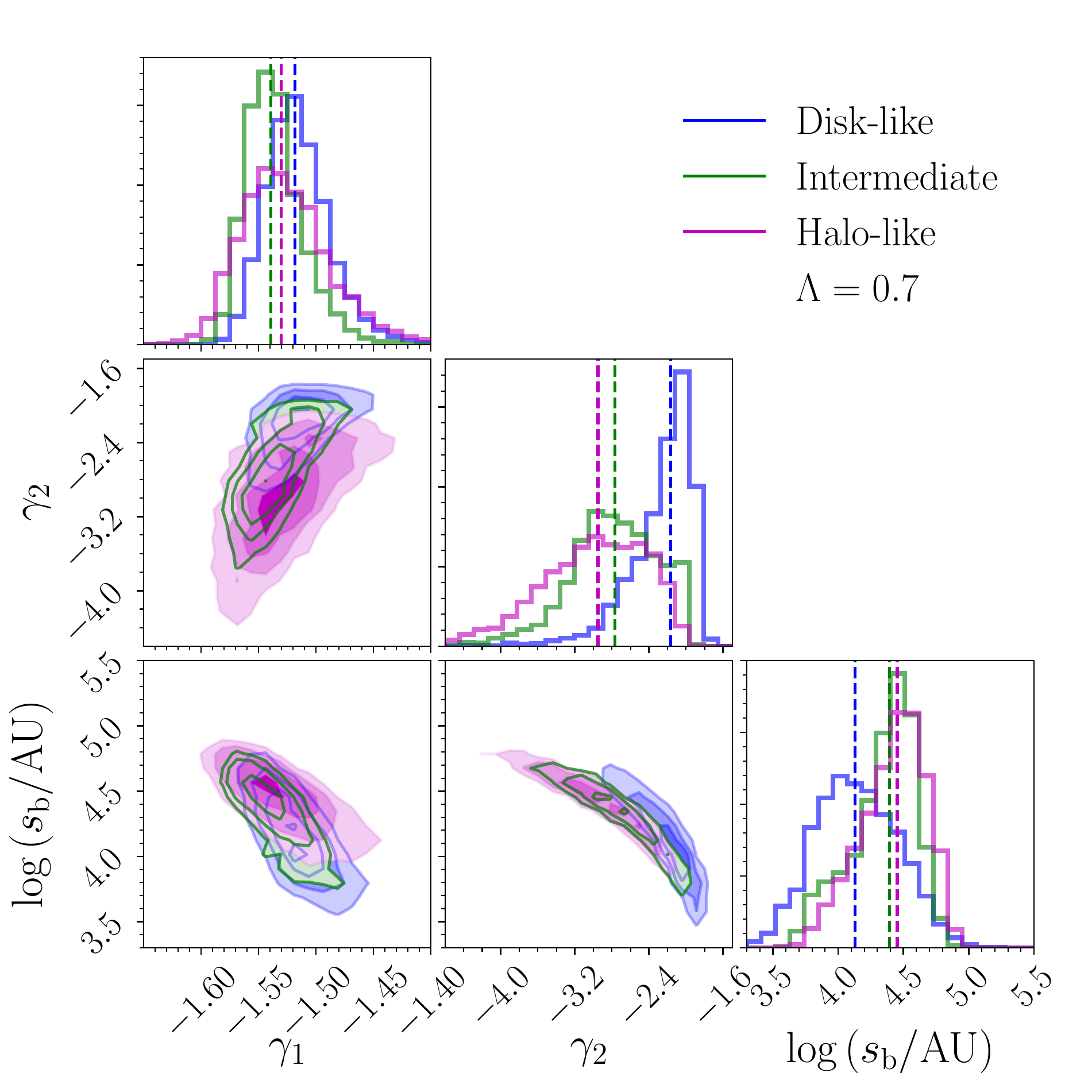}
\includegraphics[width=0.7\textwidth, trim=0.0cm 0.0cm 0.0cm 0.0cm, clip]{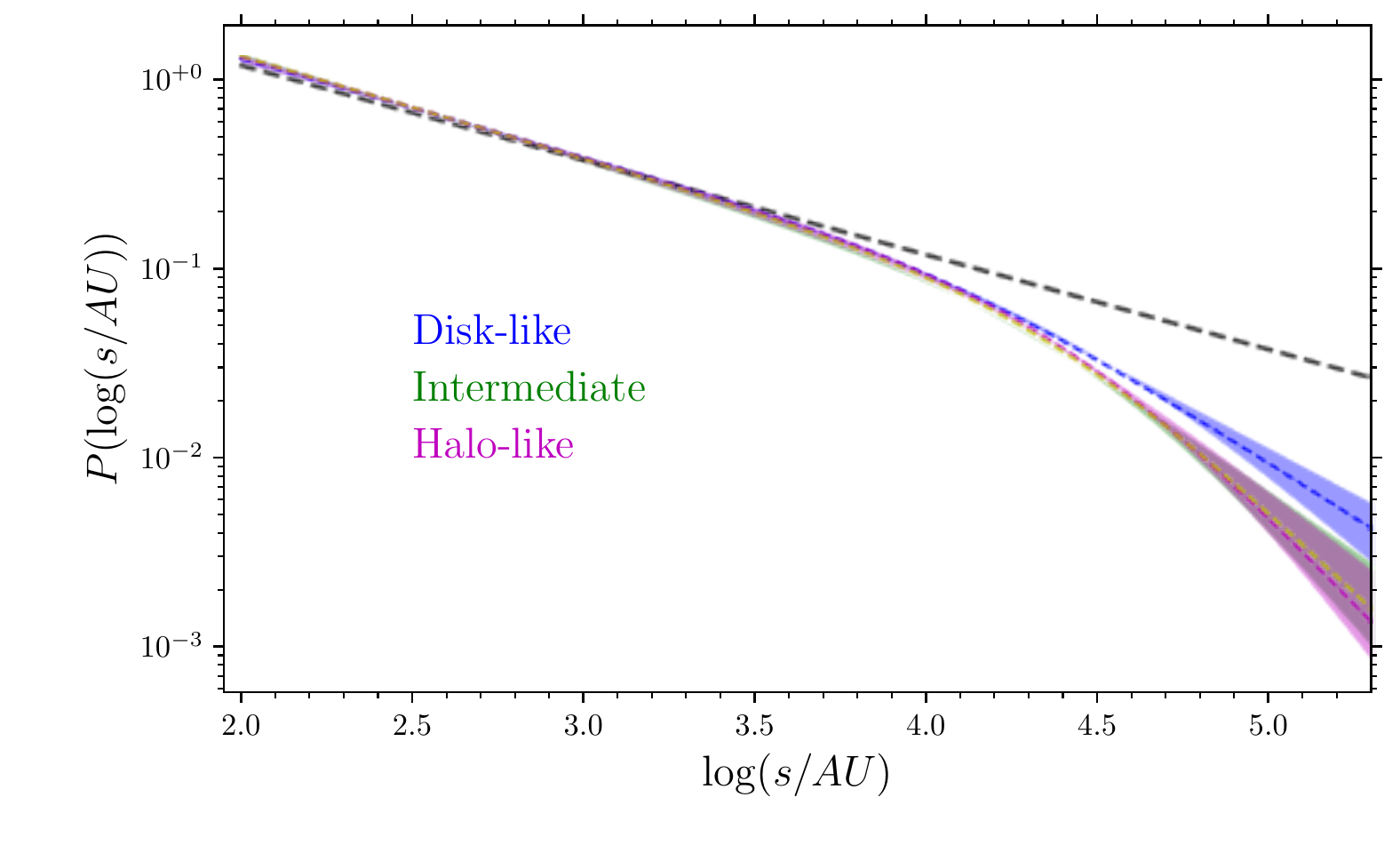}
\caption{Similar to Figure \ref{fig:fitting2}, but with a fixed smoothing parameter, i.e., $\Lambda=0.7$, and setting $s_{\rm max}=10^{5.3}$~AU. This figure highlights the wider-separation break in the separation distributions. There is a strong hint of a steeper fall-off at ultra-wide separations ($s>10^{4.5}$\,AU) for the halo-like sample than the disk sample.}\label{fig:fitting3}
\end{figure}

Figure \ref{fig:fitting3} is similar to Figure \ref{fig:fitting2},
 but with $\Lambda=0.7$ and $s_{\rm max}=10^{5.3}$~AU. The best fit values of $\vec{m} =(\gamma_1, \gamma_2, \log(s_{\rm b}/{\rm AU}))$ are
\begin{itemize}
    \item $(-1.52^{+0.03}_{-0.02},-2.17^{+0.41}_{-0.20}, 4.13^{+0.34}_{-0.31})$ for the disk-like binaries,
    \item $(-1.54^{+0.02}_{-0.03}, -2.77^{+0.45}_{-0.49}, 4.33^{+0.18}_{-0.33})$ for the intermediate binaries
    \item $(-1.53^{+0.03}_{-0.04}, -2.95^{+0.61}_{-0.54}, 4.45^{+0.21}_{-0.30})$ for the halo-like binaries. 
\end{itemize}

In the two extreme cases, we further confirm that (1) the slope for separations from $10^{2.5}$~AU to $10^{4.0}$~AU is $p(s) \approx s^{-1.5}$; (2) At $s>10^{4.0}$~AU, the separation distributions obviously deviate from the single power law; (3) At $s>10^{4.5}$~\rm AU, the slope steepens by a slightly larger amount for the old population than for the young population. 

It is worth to mention that the intermediate subsample presents bimodal parameter constraints, in particular for the parameters of $\gamma_2$ and $\log(s_{\rm b}/{\rm AU})$. The value of one peak is consistent with the value in the disk-like population, the value of the other one matches with that of the halo-like subsample, see Figure \ref{fig:fitting2} (the top panel).


\begin{thebibliography}{61}
\expandafter\ifx\csname natexlab\endcsname\relax\def\natexlab#1{#1}\fi
\bibitem[Afonso et al.(2003)]{Afonso_2003} Afonso, C., Albert, J.~N., Andersen, J., et al.\ 2003, \aap, 400, 951
\bibitem[Alcock et al.(2001)]{Alcock_2001} Alcock, C., Allsman, R.~A., Alves, D.~R., et al.\ 2001, \apjl, 550, L169
\bibitem[Allen \& Monroy-Rodr{\'\i}guez(2014)]{Allen2014ApJ} Allen, C., \& Monroy-Rodr{\'\i}guez, M.~A.\ 2014, \apj, 790, 158
\bibitem[Andrews et al.(2017)]{Andrews_2017} Andrews, J.~J., Chanam{\'e}, J., \& Ag{\"u}eros, M.~A.\ 2017, \mnras, 472, 675
\bibitem[Badenes et al.(2018)]{Badenes2018} Badenes, C., Mazzola, C., Thompson, T.~A., et al.\ 2018, \apj, 854, 147
\bibitem[Bahcall \& Soneira(1981)]{Bahcall1981} Bahcall, J.~N., \& Soneira, R.~M.\ 1981, \apj, 246, 122
\bibitem[Bahcall et al.(1985)]{Bahcall1985} Bahcall, J.~N., Hut, P., \& Tremaine, S.\ 1985, \apj, 290, 15
\bibitem[Batista et al.(2011)]{Batista2011} Batista, V., Gould, A., Dieters, S., et al.\ 2011, \aap, 529, A102
\bibitem[Binney \& Tremaine(2008)]{BT2008} Binney, J., \& Tremaine, S.\ 2008, Galactic Dynamics: Second Edition
\bibitem[Brandt(2016)]{Brandt2016} Brandt, T.~D.\ 2016, \apj, 824, L31
\bibitem[Boersma(1961)]{Boersma_1961} Boersma, J.\ 1961, \bain, 15, 291
\bibitem[Chanam{\'e} \& Gould(2004)]{CG04} Chanam{\'e}, J., \& Gould, A.\ 2004, \apj, 601, 289
\bibitem[Chanam{\'e}(2007)]{Chaname2007} Chanam{\'e}, J.\ 2007, Binary Stars as Critical Tools \& Tests in Contemporary Astrophysics, 316
\bibitem[Dhital et al.(2010)]{Dhital_2010} Dhital, S., West, A.~A., Stassun, K.~G., et al.\ 2010, \aj, 139, 2566
\bibitem[Duch{\^e}ne \& Kraus(2013)]{Duchene_2013} Duch{\^e}ne, G., \& Kraus, A.\ 2013, \araa, 51, 269
\bibitem[de Rujula et al.(1992)]{DJM92} de Rujula, A., Jetzer, P., \& Masso, E.\ 1992, \aap, 254, 99
\bibitem[El-Badry \& Rix(2018)]{El-Badry2018a} El-Badry, K., \& Rix, H.-W.\ 2018, \mnras, 480, 4884
\bibitem[El-Badry \& Rix(2019)]{EBR2019} El-Badry, K., \& Rix, H.-W.\ 2019, \mnras, 482, L139
\bibitem[El-Badry et al.(2019)]{EBR2019b} El-Badry, K., Rix, H.-W., Tian, H., et al.\ 2019, \mnras, 489, 5822
\bibitem[Evans et al.(2018)]{Evans2018} Evans, D.~W., Riello, M., De Angeli, F., et al.\ 2018, \aap, 616, A4

\bibitem[Fouesneau et al.(2019)]{Fouesneau2019} Fouesneau, M., Rix, H.-W., von Hippel, T., et al.\ 2019, \apj, 870, 9
 \bibitem[Foreman-Mackey(2016)]{corner} Foreman-Mackey, D.\ 2016, The Journal of Open Source Software, 1, 24
\bibitem[Foreman-Mackey et al.(2013)]{FormanMackey_2013} Foreman-Mackey, D., Hogg, D.~W., Lang, D., et al.\ 2013, \pasp, 125, 306

\bibitem[Gaia Collaboration et~al. (2016)]{gaia2016}{Gaia Collaboration}, {Brown}, A. G. A., {Vallenari}, A., {Makarov}, V. V., {et~al.} 2016, A\&A, 595, A2
\bibitem[Gaia Collaboration et~al. (2018)]{gaia2018}{Gaia Collaboration:} {Brown}, A. G. A., {Vallenari}, A., {Makarov}, V. V., {et~al.} 2018, A\&A, 616, A1
\bibitem[Green(2016)]{Green_2016} Green, A.~M.\ 2016, \prd, 94, 063530 
\bibitem[Kouwenhoven et al.(2010)]{Kouwenhoven2010} Kouwenhoven, M.~B.~N., Goodwin, S.~P., Parker, R.~J., et al.\ 2010, \mnras, 404, 1835
\bibitem[L{\'e}pine, \& Bongiorno(2007)]{Lepine_2007} L{\'e}pine, S., \& Bongiorno, B.\ 2007, \aj, 133, 889
\bibitem[Liu(2019)]{liu2019} Liu, C.\ 2019, arXiv e-prints, arXiv:1907.02250
\bibitem[Lindegren et al.(2018)]{Lindegren2018} Lindegren, L., Hern{\'a}ndez, J., Bombrun, A., et al.\ 2018, \aap, 616, A2
\bibitem[Moe, \& Di Stefano(2017)]{MD2017} Moe, M., \& Di Stefano, R.\ 2017, \apjs, 230, 15
\bibitem[Moe \& Kratter(2018)]{MK2018} Moe, M., \& Kratter, K.~M.\ 2018, \apj, 854, 44
\bibitem[Moeckel \& Bate(2010)]{Moeckel2010} Moeckel, N., \& Bate, M.~R.\ 2010, \mnras, 404, 721 
\bibitem[Moeckel \& Clarke(2011)]{Moeckel2011} Moeckel, N., \& Clarke, C.~J.\ 2011, \mnras, 415, 1179 
\bibitem[Oelkers et al.(2017)]{Oelkers_2017} Oelkers, R.~J., Stassun, K.~G., \& Dhital, S.\ 2017, \aj, 153, 259
\bibitem[Pittordis, \& Sutherland(2019)]{Pittordis_2019} Pittordis, C., \& Sutherland, W.\ 2019, \mnras, 1859
\bibitem[Quinn et al.(2009)]{Quinn09} Quinn, D.~P., Wilkinson, M.~I., Irwin, M.~J., et al.\ 2009, \mnras, 396, L11
\bibitem[Raghavan et al.(2010)]{Raghavan2010} Raghavan, D., McAlister, H.~A., Henry, T.~J., et al.\ 2010, \apjs, 190, 1
\bibitem[Reipurth, \& Mikkola(2012)]{Reipurth_2012} Reipurth, B., \& Mikkola, S.\ 2012, \nat, 492, 221
\bibitem[Retterer \& King(1982)]{Retterer1982} Retterer, J.~M., \& King, I.~R.\ 1982, \apj, 254, 214 
\bibitem[Robin et al.(2003)]{Robin2003} Robin, A.~C., Reyl{\'e}, C., Derri{\`e}re, S., et al.\ 2003, \aap, 409, 523
\bibitem[Rybizki et al.(2018)]{Rybizki2018} Rybizki, J., Demleitner, M., Fouesneau, M., et al.\ 2018, \pasp, 130, 74101
\bibitem[Jiang, \& Tremaine(2010)]{Jiang_2010} Jiang, Y.-F., \& Tremaine, S.\ 2010, \mnras, 401, 977
\bibitem[Savedoff(1966)]{Savedoff_1966} Savedoff, M.~P.\ 1966, \aj, 71, 396
\bibitem[Sharma et al.(2011)]{Sharma2011} Sharma, S., Bland-Hawthorn, J., Johnston, K.~V., et al.\ 2011, \apj, 730, 3
\bibitem[Sch\"onrich (2012)]{schonrich2012} Sch\"onrich, R.\ 2012 \mnras, 427, 274 
\bibitem[Tian et al.(2015)]{tian2015} Tian, H.-J., Liu, C., Carlin, J.~L., et al.\ 2015, \apj, 809, 145
\bibitem[Tisserand et al.(2007)]{Tisserand07} Tisserand, P., Le Guillou, L., Afonso, C., et al.\ 2007, \aap, 469, 387
\bibitem[van Albada(1968)]{van1968} van Albada, T.~S.\ 1968, \bain, 20, 57 
\bibitem[Wyrzykowski et al.(2008)]{Wyrzykowski08} Wyrzykowski, L., Kozlowski, S., Belokurov, V., et al.\ 2008, Manchester Microlensing Conference, 11
\bibitem[Wasserman, \& Weinberg(1987)]{1987ApJ...312..390W} Wasserman, I., \& Weinberg, M.~D.\ 1987, \apj, 312, 390
\bibitem[Weinberg et al.(1987)]{Weinberg_1987} Weinberg, M.~D., Shapiro, S.~L., \& Wasserman, I.\ 1987, \apj, 312, 367
\bibitem[Widmark et al.(2018)]{Widmark18} Widmark, A., Leistedt, B., \& Hogg, D.~W.\ 2018, \apj, 857, 114
\bibitem[Yoo et al.(2004)]{Y04} Yoo, J., Chanam{\'e}, J., \& Gould, A.\ 2004, \apj, 601, 311
\bibitem[Zapatero Osorio, \& Mart{\'\i}n(2004)]{Zapatero2004} Zapatero Osorio, M.~R., \& Mart{\'\i}n, E.~L.\ 2004, \aap, 419, 167

\end{thebibliography}
\end{document}